\documentclass{jpp}
\usepackage{amsmath}
\usepackage{amssymb}
\usepackage{xcolor}
\usepackage{graphicx}
\usepackage{subcaption}
\usepackage{siunitx}                                
\usepackage{overpic}                                
\usepackage{booktabs}                               
\usepackage{multirow}
\usepackage{hyperref}
\usepackage{upgreek}
\usepackage{comment}
\newcommand{\Ire}{{I_{\rm re}}}
\newcommand{\Iremax}{{I_{\rm re}^{\rm max}}}
\newcommand{\Irenf}{{I_{\rm re}^{95\,\%}}}
\newcommand{\Iohm}{{I_\Upomega}}
\newcommand{\johm}{{j_\Upomega}}

\newcommand{\Iohmfin}{{I_\Upomega^{\rm final}}}
\newcommand{\Ip}{{I_{\rm p}}}
\newcommand{\Ipin}{{I_{\rm p}^{t=0}}}
\newcommand{\taucq}{{\tau_{\Upomega,{\rm CQ}}}}
\newcommand{\etac}{{\eta_{\rm tr}}}
\newcommand{\nT}{{n_{\rm T}}}
\newcommand{\nD}{{n_{\rm D}}}
\newcommand{\nHe}{{n_{\rm He}}}
\newcommand{\nNe}{{n_{\rm Ne}}}
\newcommand{\nAr}{{n_{\rm Ar}}}

\newcommand{\dBB}{{\delta B/B}}

\newcommand{\nth}{{n_{\rm th}}}

\newcommand{\nst}{{n_{\rm st}}}
\newcommand{\nre}{{n_{\rm re}}}
\newcommand{\fst}{{f_{\rm st}}}
\newcommand{\Tth}{{T_{\rm th}}}
\newcommand{\me}{{m_{\rm e}}}
\newcommand{\Lcf}{{\mathcal{L}}}
\newcommand{\DREAM}{\textsc{Dream}}

\newcommand{\deuterium}{D}
\newcommand{\tritium}{T}
\newcommand{\deuteriumtritium}{DT}

\newcommand{\helium}{He}
\newcommand{\neon}{Ne}
\newcommand{\argon}{Ar}

\newcommand{\dd}{{\rm d}}
\newcommand{\pdv}[2]{\frac{\partial #1}{\partial #2}}
\newcommand{\dv}[2]{\frac{\dd #1}{\dd #2}}
\newcommand{\abs}[1]{\left| #1 \right|}

\shorttitle{Runaways in SPARC disruptions}
\shortauthor{I.~Ekmark et al}

\title{Runaway electron generation in disruptions mitigated by deuterium and noble gas injection in SPARC} 

\author{I.~Ekmark\aff{1}
    \corresp{\email{ida.ekmark@chalmers.se}}, M.~Hoppe\aff{2}, R.~A.~Tinguely\aff{3}, R.~Sweeney\aff{4}, 
    T.~Fülöp\aff{1}, \and I.~Pusztai\aff{1}}

\affiliation{
\aff{1}Department of Physics, Chalmers University of Technology, G\"{o}teborg, SE-41296, Sweden
\aff{2} Department of Electrical Engineering, KTH Royal Institute of Technology, Stockholm, SE-11428, Sweden
\aff{3}Plasma Science and Fusion Center, Massachusetts Institute of Technology, Cambridge, MA 01239, USA
\aff{4}Commonwealth Fusion Systems, Devens, MA, USA
}

\begin{document}

\maketitle

\begin{abstract}
One of the critical challenges in future high current tokamaks is the avoidance of runaway electrons during 
disruptions. 
Here, we investigate disruptions mitigated with combined deuterium and noble gas injection in SPARC. 
We use multi-objective Bayesian optimization of the densities of the injected material, taking into account 
limits on the maximum runaway current, the transported fraction of the heat loss, and the current quench 
time. 
The simulations are conducted using the numerical framework \DREAM{} (Disruption Runaway Electron Analysis 
Model). 
We show that during deuterium operation, runaway generation can be avoided with material injection, even 
when we account for runaway electron generation from DD-induced Compton scattering. 
However, when including the latter, the region in the injected-material-density space corresponding to 
successful mitigation is reduced. 
During deuterium--tritium operation, acceptable levels of runaway current and transported heat losses are 
only obtainable at the highest levels of achievable injected deuterium densities. 
Furthermore, disruption mitigation is found to be more favourable when combining deuterium with neon, 
compared to deuterium combined with helium or argon.
\end{abstract}

\section{Introduction}
One of the main obstacles to the successful operation of tokamak fusion reactors is plasma-terminating 
disruptions, caused by instabilities that lead to a sudden loss of plasma confinement. 
During such events, the plasma facing components can be damaged by high heat loads, and electromagnetic 
forces can cause significant mechanical stress on the machine \citep{hollmann2015}. 
Disruptions are characterized by a rapid drop in plasma temperature (thermal quench (TQ)), causing a 
subsequent, slower decrease in plasma current due to increased resistivity (current quench (CQ)), which in 
turn induces an electric field capable of accelerating electrons to relativistic speeds 
\citep{helander2002, breizman2019}. 
One of the major challenges associated with disruptions is the generation of highly energetic runaway 
electron (RE) beams.
If control of such a RE beam is lost, a large fraction of the stored energy can be deposited onto the first 
wall, sometimes in a highly localized impact area \citep{jepu2024}. 

This challenge is especially relevant for future devices such as SPARC and ITER, as the dominating RE 
generation mechanism, avalanche multiplication, is exponentially sensitive to the initial plasma current 
\citep{rosenbluth1997}, which at full power will be \SI{8.7}{MA} and \SI{15}{MA} for SPARC and ITER, 
respectively.
There are several proposed mitigation methods, most prominently massive material injection (MMI) of a 
combination of radiating impurities (e.g. \neon{} or \argon) and hydrogen isotopes, either in gaseous or 
solid state.
In ITER, the reference concept for the disruption mitigation system is shattered pellet injection (SPI)
\citep{baylor2019}. 
SPARC, on the other hand, will first deploy a simpler massive gas injection (MGI) system, and will 
additionally utilize a passive conducting coil with 3D structure -- the runaway electron mitigation coil 
(REMC) -- acting to deconfine the REs faster than they are generated \citep{sweeney2020}.

RE mitigation in SPARC using MGI, both with and without the REMC, has been studied in previous works 
\citep{tinguely2021, izzo2022, tinguely2023}, and the addition of the REMC is shown to decrease the runaway 
current by several MAs. 
Runaway dynamics in ITER disruptions with MMI (both MGI and SPI) has been extensively studied by 
\citep{vallhagen2020,vallhagen2024}. 
Similar studies have been performed for STEP (Spherical Tokamak for Energy Production) disruptions 
\citep{berger2022,fil2024}.
In other works \citep{pusztai2023, ekmark2024}, Bayesian optimization was utilized to explore a large range 
of MMI density combinations for \deuterium{} and \neon{} in ITER. 
In addition to injected density magnitudes, \cite{pusztai2023} studied optimal radial distributions of the 
injected material. 
The main focus of the optimization was to minimize the runaway current, but these works also considered 
figures of merits corresponding to heat loads from plasma particle transport and mechanical stresses from 
electromagnetic forces. 
None of these investigations found parameter regions with successful disruption mitigation for 
\deuteriumtritium{} operation in ITER.
However, there has been no such comprehensive exploration of the MMI injection density space for the 
compact, high field device SPARC. 

In this paper, we  explore the space of injected \deuterium{} and noble gas densities in SPARC, without the 
effect of the REMC. 
Since simulations indicate that the REMC is capable of significantly mitigating the generation of a runaway 
current, we here aim to study disruption mitigation in SPARC without the REMC. 
This will isolate the mitigation effect of the MMI and give a conservative view of the success of the RE 
mitigation. 
The aim is to find parameter regions of favourable disruption mitigation and successful RE avoidance. 
Sample selection is done utilizing Bayesian optimization and the simulations are performed using the 
simulation tool \DREAM{} \citep{dream}. 
More specifically, we employ the pitch-angle averaged kinetic model for superthermal electrons in \DREAM, 
which accurately captures the seed generation mechanisms \citep{ekmark2024}, yet it is sufficiently 
computationally inexpensive to allow numerical optimization over a large parameter space.

\section{Simulation setup and plasma model}
For the optimization of disruption mitigation in SPARC, we have simulated disruption scenarios using the 
\DREAM{} code, developed especially for accurate and efficient studies of REs in fusion plasmas.
In \S\ref{sec:sparc}, we describe how the SPARC disruptions have been modelled, while the electron and 
plasma parameter evolution is detailed in \S\ref{sec:electron}, and the optimization specifics are 
presented in \S\ref{sec:opt}.

\subsection{SPARC disruption scenario}\label{sec:sparc}
\begin{figure}
    \centering
    \begin{subfigure}[t]{4.92cm}
        \begin{overpic}[width=4.92cm]{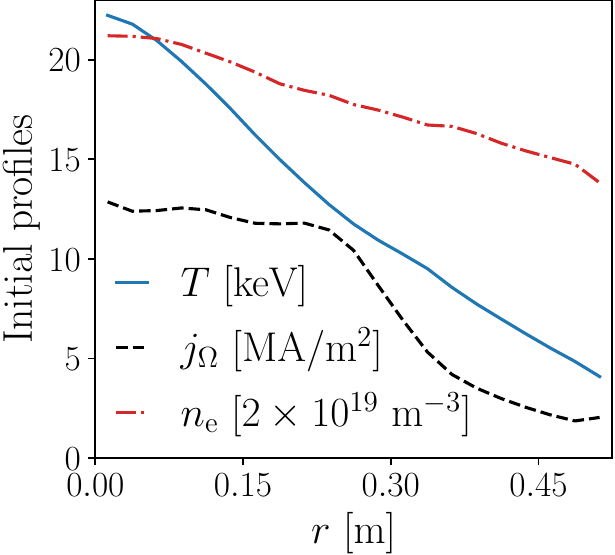}
            \put(89, 82){(a)}
        \end{overpic}
        \phantomcaption
        \label{fig:set_IC}
    \end{subfigure}
    \hspace{0.005cm}
    \begin{subfigure}[t]{2.955cm}
        \begin{overpic}[width=2.955cm]{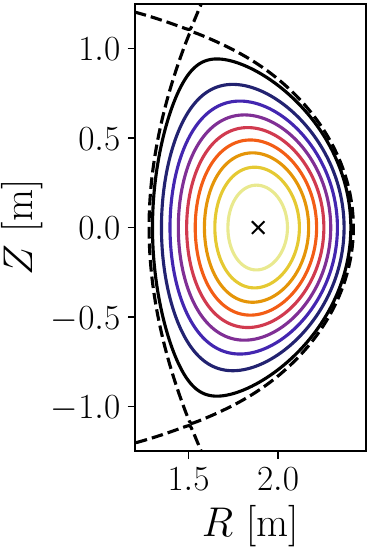}
            \put(55, 91){(b)}
        \end{overpic}
        \phantomcaption
        \label{fig:set_eq}
    \end{subfigure}
    \hfill
    \begin{subfigure}[t]{4.92cm}
        \begin{overpic}[width=4.92cm]{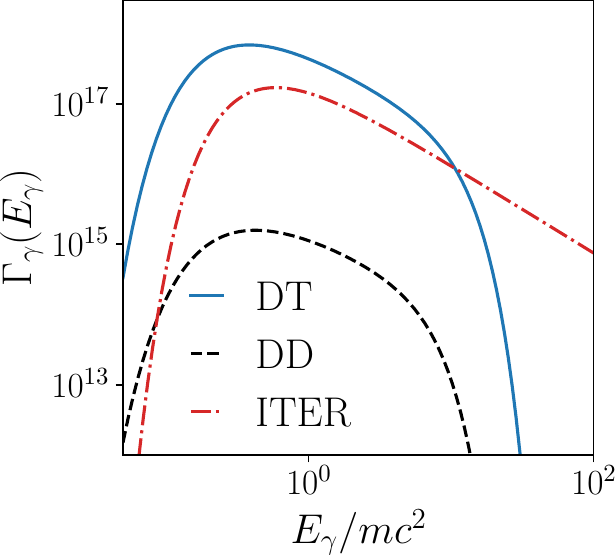}
            \put(21, 82){(c)}
        \end{overpic}
        \phantomcaption
        \label{fig:set_C}
    \end{subfigure}
    \caption{\small Characteristics of the SPARC primary reference discharge. 
    (a) Initial plasma temperature (solid blue), density (dash-dotted red), and Ohmic 
    current density (dashed black) profiles.
    (b) Equilibrium flux surfaces (solid), plasma separatrix (dashed) and magnetic axis (cross). 
    (c) Photon flux energy spectrum for a DT-plasma in SPARC with a total photon flux of 
    \SI{1.4e18}{\per\meter\squared\per\second} (solid blue) and for a DD-plasma in SPARC with a total photon 
    flux of \SI{3.3e15}{\per\meter\squared\per\second} (dashed black) compared to a DT-plasma in ITER with a 
    total photon flux of \SI{e18}{\per\meter\squared\per\second} (dash-dotted red). }
    \label{fig:set}
\end{figure}

We have initialized our disruption simulations with parameters corresponding to the SPARC primary reference 
discharge \citep{creely2020}, which has a plasma current ${\Ip=\SI{8.7}{MA}}$. 
The primary reference discharge is a \deuteriumtritium{} (50–50 isotope mix) plasma, but we have also 
considered a pure \deuterium{} plasma with the same initial conditions. 
Notably, the assumptions of identical initial conditions for pure \deuterium{} and \deuteriumtritium{} 
plasmas is somewhat simplistic, but enables us to compare the effects of Compton scattering and tritium 
beta decay on the runaway dynamics more easily. 
The initial plasma temperature, density and Ohmic current profiles from 
\citep{RodriguezFernandez2020,RodriguezFernandez2022} used for the simulations are shown in 
figure~\ref{fig:set_IC}. 
The electron and ion temperatures are initialized identically, but evolved separately. 
Furthermore, the equilibrium used has the on axis magnetic field ${B_0=\SI{12.5}{T}}$\footnote{Note that 
this magnetic field includes both the vacuum toroidal field of \SI{12.2}{T} and the magnetic field 
contribution from the plasma. }, 
major radius ${R_0=\SI{1.89}{m}}$ taken at the magnetic axis location, and minor radius ${a=\SI{0.525}
{m}}$, which have been obtained from FREEGS equilibrium simulations 
\citep{RodriguezFernandez2020,RodriguezFernandez2022}. 
In figure~\ref{fig:set_eq}, the closed flux surfaces from the equilibrium simulations are plotted. 
The wall radius, which is a parameter in the boundary condition for Ampère's law, was set to 
${b=\SI{0.621}{m}}$ to match the available poloidal magnetic energy ${E_{\rm mag}=\SI{52.3}{MJ}}$ 
calculated using COMSOL by \citet{tinguely2021}, and the resistive wall time was set to \SI{80}{ms} 
\citep{battey2024}.
As noted by \citet{tinguely2021}, the choice of wall radius does have an impact on the RE dynamics, but 
the qualitative optimization results are not affected to the same degree, according to \citep{ekmark2024}. 
The evolution of the magnetic flux surface shaping and current density relaxation \citep{Pusztai_relax} 
during the TQ are not taken into account in the simulations. 

The disruption simulation is started with the deposition of \deuterium{} and noble gas (either \helium, 
\neon{} or \argon), instantly and uniformly across the plasma volume. 
During the disruption, the plasma heat losses are caused by convective losses of energetic electrons, heat 
conduction of the bulk electrons, as well as radiation which is more efficient at lower (below 
$\sim100$-$\SI{1000}{eV}$) temperatures. 
The magnetic field stochastization during the TQ cannot be modelled self-consistently in \DREAM. 
Instead, we employ spatially and temporally constant magnetic perturbations with a relative amplitude of 
$\dBB=\SI{0.3}{\percent}$, which was chosen to achieve a TQ duration of $\sim0.1$-\SI{1}{ms} 
\citep{sweeney2020} (the TQ time is presented for a few sample simulations in table~\ref{tab:cases}). 
Note that the enforced transport due to magnetic perturbations is initialized simultaneously with the 
material deposition. 
The perturbations are active until the end of the TQ, which we define by the mean temperature falling 
below \SI{20}{\electronvolt}. 

During the TQ the transport of superthermal electrons, REs and heat is calculated with the 
Rechester--Rosenbluth model (\citeyear{RechesterRosenbluth1978}), with the diffusion coefficient 
${D^{rr}\propto \abs{v_\parallel} R_0(\dBB)^2}$. 
For the heat transport, $\abs{v_\parallel}$ is replaced by the electron thermal speed and for the RE 
transport by the speed of light. 
The Rechester--Rosenbluth model, however, does not account for the energy and angular dependence of the RE 
distribution, or for finite Larmor radius and orbit width effects \citep{sarkimaki2020}, but it gives an 
upper bound on the effect of the RE transport, as noted by \citet{svensson2021} and \citet{pusztai2023}.

Lower values of $\dBB$ reduce both the transported heat and RE losses. 
Yet the reduction in transported heat also slows the TQ temperature decay, thereby decreasing the hot-tail 
RE seed generation. 
Thus, the net effect of decreasing $\dBB$ is, somewhat counter-intuitively, that the region of ``safe'' MMI 
space increases.
Simulations with a TQ duration $>\SI{3}{ms}$ are deemed to have an incomplete TQ, meaning that the amount 
of material that has been injected is not enough to cause a proper disruption. 
For such scenarios we do not run the CQ simulation, and instead penalize the sample with a high cost 
function value.

As the magnetic flux surfaces are expected to heal after the TQ, the transport of superthermal electrons 
and REs is switched off during the CQ; however, a small heat transport, with $\dBB=\SI{0.04}{\percent}$ in 
accordance with previous works \citep{pusztai2023, ekmark2024}, is retained to suppress the development of 
non-physical hot Ohmic current channels \citep{putvinski1997,feher2011}. 
This transport level is orders of magnitude lower than the radiative heat losses at post-TQ temperatures, 
so it does not influence the evolution of the temperature and other plasma parameters.

\subsection{Electron dynamics and runaway modelling in DREAM}\label{sec:electron}
Electrons in \DREAM{} \citep{dream} are divided into three populations -- the thermal, superthermal and 
runaway populations -- based on their momentum, and they are generally evolved using different models. 
Importantly, the superthermal electrons, with momenta ${p\sim 0.01}$ -- $1\ \me c$ (with $\me$ being the 
electron rest mass) of the same order of magnitude as the typical values for the critical momentum 
for RE generation, can be simulated using a kinetic model which has been analytically averaged over 
pitch-angle. 
Note that this pitch-angle averaged treatment is an accurate approximation at these moderate momenta, as 
the pitch-angle scattering is sufficiently strong to approximately isotropize the electron distribution. 
In this work, we have used this reduced kinetic model for the superthermal electrons, while the electrons 
in the thermal and runaway populations are modelled as fluids. 

The advantage of this approach is that it greatly reduces the computational cost compared to fully kinetic 
simulations, while still enabling the runaway seed generation to be accurately simulated. 
Since the kinetic modelling of the seed generation entails more relaxed assumptions, it is typically more 
accurate than fluid modelling, especially in simulations with rapidly varying plasma parameters. 
The differences between fluid and kinetic treatment of the seed generation mechanisms can be quite 
significant, as shown in \citep{ekmark2024}.

The superthermal electrons will be described by their distribution function, from which relevant velocity 
moments, or fluid quantities, can be obtained, e.g. the superthermal electron density $\nst$ and current 
density $j_{\rm st}$. 
The distribution function, $\fst$, will be evolved according to the bounce-averaged kinetic equation, 
derived in appendix B.2. in \citep{dream} and referred to as the ``isotropic'' model,
\begin{align}
    \label{eq:dfdt_AD} \pdv{\fst}{t}&=
    \frac{1}{p^2}\pdv{}{p} \left[ p^2\left( 
        - A^p \fst + \hat{D}^{pp} \pdv{\fst}{p} 
    \right) \right]
    +\frac{1}{V^\prime}\pdv{}{r}\left[ V^\prime D^{rr} \pdv{\fst}{r} \right]+ S_{\rm T}+ S_{\rm C},
\end{align}
where $V^\prime$ is the spatial Jacobian, and collisions are modelled by a test-particle Fokker-Planck 
operator. 
Note that the momentum $p$ is normalized to $\me c$. 
The distribution function is evolved for $p\in[0,\, p_{\rm re}]$, where we have chosen the upper momentum 
boundary as $p_{\rm re}=2.5\,\me c$, and $\fst(p> p_{\rm re})=0$.
In \eqref{eq:dfdt_AD}, the momentum advection term $A^p$ consists of slowing down due to the bremsstrahlung 
and collisions.
Furthermore, $\hat{D}^{pp}=D^{pp}+\mathcal{D}_E$, where $D^{pp}$ is the diffusive component of the 
collision operator, and ${\mathcal{D}_E\propto(e\boldsymbol{E}\cdot \boldsymbol{B})^2/( B^2 \nu_{\rm D})}$ 
describes the effect of the electric field in the presence of strong pitch-angle scattering, with the 
pitch-angle scattering collision frequency $\nu_{\rm D}=\nu_{\rm D}(p)$. 
The radial diffusion $D^{rr}$ is evaluated using the Rechester--Rosenbluth model 
(\citeyear{RechesterRosenbluth1978}).

The source term 
\begin{align}
    S_{\rm T} &=\label{eq:dfdt_T}
    C\frac{\ln2}{4\pi} \frac{\nT}{\tau_{\rm T}}\frac{1}{p^2} 
        \frac{p \gamma \left(\gamma_{\rm max}-\gamma\right)^2}{1-\exp(-4\pi\alpha/\beta)}
            \Theta(p_{\rm max} - p),
\end{align}
describes the generation of superthermal electrons from tritium beta decay, derived by \citet{ekmark2024}.
In \eqref{eq:dfdt_T}, the fine structure constant $\alpha\approx1/137$ and $C$ is a normalization constant 
used to ensure that the integrated source yields the \tritium{} decay rate $(\ln 2)\, \nT/\tau_{\rm T}$, 
where $\tau_{\rm T}\approx 4500 $ days is the half-life for \tritium{} and $\nT$ is the \tritium{} density. 
Additionally, $\gamma=\sqrt{p^2+1}$ is the Lorentz factor and $\beta=p/\gamma$ is the normalized speed. 
During tritium beta decay, electrons will be generated with a kinetic energy $W\in[0,\ W_{\rm max}]$, 
with $W_{\rm max}=\SI{18.6}{keV}$ determining $p_{\rm max}$ and $\gamma_{\rm max}$.

The generation of superthermal electrons from Compton scattering due to photons emitted from the walls 
during activated operation is described by \citep{ekmark2024}
\begin{align}
    S_{\rm C} &=\label{eq:dfdt_C}
    \frac{n_{\rm e,tot}}{2}\frac{1}{p^2}\int_{W_{\gamma0}}^\infty \Gamma_\gamma(W_\gamma)
        \dv{\sigma}{\Omega} \frac{\beta}{\left(\frac{W_\gamma}{\me c^2}+1-\gamma\right)^2}\dd W_\gamma,
\end{align}
where $n_{\rm e,tot}$ is the total electron density in the plasma and $\dd \sigma / \dd \Omega$ is the 
Klein-Nishina differential cross section \citep{kleinNishina}.
Here, the photon flux energy spectrum 
\begin{equation}
    \Gamma_\gamma(W_\gamma)
    =\Gamma_{\rm flux}\exp[-\exp(-z)-z+1]\left/\int \exp[-\exp(-z)-z+1]\, \dd W_\gamma\right.,
\end{equation}
where $z = [\ln(W_\gamma\ [\si{MeV}])+C_1]/C_2+C_3(W_\gamma\ [\si{MeV}])^2$, is based on the functional 
form used for ITER by \cite{solis2017}.
The $W_\gamma^2$-term in the expression for $z$ is added to the form used by \cite{solis2017} to achieve a 
steeper decrease for higher photon energies, in accordance with the photon energy spectrum data used for 
the fit of the parameters $C_1$, $C_2$ and $C_3$.

The data for the photon energy spectrum are preliminary results obtained from Monte Carlo N-Particle 
Transport (MCNP) calculations by Commonwealth Fusion Systems, and the photon energy spectrum is plotted 
in figure~\ref{fig:set_C} together with the ITER spectrum. 
Note that despite SPARC having a lower fusion power than ITER, its much smaller size causes the total 
photon fluxes to be similar -- for both, ${\Gamma_{\rm flux}\sim\SI{e18}{\per\meter\squared\per\second}}$. 
The parameters used to represent the photon flux spectrum for both \deuteriumtritium{} and \deuterium{} 
operation are presented in table~\ref{tab:compton}, together with the corresponding ITER parameters. 
We have assumed that the flux from the activated walls reduces by a factor of $10^4$ after the TQ when the 
fusion reactions, and corresponding neutron generation, cease. 


\begin{table}
	\centering
	\caption{Total photon flux and fitted photon flux spectrum parameters used for the source term for
    energetic electrons generated by Compton scattering.
    The corresponding photon flux spectra are plotted in figure \ref{fig:set_C}.
    The photon flux spectrum parameters were obtained by fitting data from MCNP calculations.}
    \label{tab:compton}
    \begin{tabular}{l c c c c}
		\toprule
                 & $\Gamma_{\rm flux}\ [1/(\rm m^2 s)]$ & $C_1$   & $C_2$   & $C_3$   \\
        \midrule
        SPARC DT & $1.4\times10^{18}$                   & $1.525$ & $0.850$ & $0.038$ \\
        SPARC DD & $3.3\times10^{15}$                   & $1.627$ & $0.919$ & $0.094$ \\
        ITER DT  & $1\times10^{18}$                     & $1.2$   & $0.8$   & $0.0$   \\
	\end{tabular}
\end{table}

The runaway electrons will be evolved through their particle density $\nre$, according to \citep{dream}
\begin{subequations}
\begin{align}
    \pdv{\nre}{t}&=\Gamma_{\rm ava}\nre+\phi_{\rm st}^p+\gamma_{\rm C}
                        +\frac{1}{V^\prime}\pdv{}{r}\left[ V^\prime D^{rr} \pdv{\nre}{r} \right],\\
    \phi_{\rm st}^p&=   4\pi V' p_{\rm re}^2\left[ 
                            - A^p \fst + \hat{D}^{pp}\pdv{\fst}{p} 
                        \right]_{p=p_{\rm re}},\\
    \gamma_{\rm C}&=\int_{p>p_{\rm re}}S_{\rm C}\ \dd^3\boldsymbol{p}.
\end{align}
\end{subequations}
Here, the avalanche growth rate $\Gamma_{\rm ava}$ is determined by the model of \citet{hesslow2019}, and 
the radial transport by the Rechester--Rosenbluth model (\citeyear{RechesterRosenbluth1978}) through 
$D^{rr}$.
The flux of particles $\phi_{\rm st}^p$ from the superthermal population is determined by the momentum 
space flux through the upper boundary $p_{\rm re}$ of the superthermal grid, on which the superthermal 
distribution function is defined. 
This momentum space flux naturally includes the RE source generation categorized as Dreicer generation and 
hot-tail generation, as well as the generation of REs from Compton scattering and tritium beta decay with 
$p<p_{\rm re}$. 
Compton scattering will also generate REs with $p>p_{\rm re}$, which is accounted for with 
$\gamma_{\rm C}$. 
Note that the same is not true for tritium beta decay, as we have chosen $p_{\rm re}$ such that 
$p_{\rm max}<p_{\rm re}$. 
The REs will be assumed to travel at the speed of light, which is typically valid in reactor-scale tokamak 
disruptions, yielding the current density $j_{\rm re}=\nre e c$ as supported by \cite{buchholz2023}. 

The thermal electrons will be represented by their density $\nth$, temperature $\Tth$ and the Ohmic 
current density $\johm$. 
Quasineutrality is used to constrain the thermal density, i.e. ${\nth=n_{\rm free} - \nst - \nre}$, while 
the Ohmic current density is evolved through ${\johm=\sigma (\boldsymbol{E}\cdot \boldsymbol{B})/B}$. 
For the electrical conductivity $\sigma$, we use the Sauter-Redl formula \citep{redl2021}, which accounts 
for trapping effects. 
The temperature $\Tth$ is evolved through the thermal energy $W_{\rm th}=3\nth\Tth/2$, according to 
\citep{dream}
\begin{equation}
    \label{eq:Wth}
    \pdv{W_{\rm th}}{t}=
    \frac{\johm}{B}\boldsymbol{E}\cdot \boldsymbol{B}-\nth\sum_i\sum_{j=0}^{Z_i-1}n_i^{(j)}L_i^{(j)} 
    + Q_{\rm c} + \frac{1}{V^\prime}\pdv{}{r}\left[ V^\prime D^{rr}\frac{3\nth}2\pdv{\Tth}{r}\right], 
\end{equation}
and outside of the plasma $\Tth(r>a)=0$. 
The first term of \eqref{eq:Wth} models the Ohmic heating and the third term heating from collisions with ions and 
non-thermal electrons, while the fourth accounts for the radial transport. 
Energy loss due to inelastic atomic processes are accounted for in the second term for each atomic species 
$i$ with atomic number $Z_i$, namely line-, recombination- and bremsstrahlung radiation, as well as changes 
in potential energy due to ionization and recombination, all encompassed by the effective rate coefficient 
$L_i^{(j)}$. 
The ionization, recombination and radiation rates are taken from the ADAS database for the noble gases. 
For the hydrogen isotopes, data from the AMJUEL database \citep{amjuel} is used instead, to account for 
opacity to Lyman radiation.
This has been shown to be important at high \deuterium{} densities \citep{vallhagen2020,vallhagen2022}.

\subsection{Disruption optimization for SPARC}\label{sec:opt}
As figures of merit for quantifying the efficiency of the disruption mitigation, we use a combination of 
the runaway current $\Ire$, fraction of heat transported to the plasma facing components $\etac$ (including 
heat loss through the convective losses of superthermal electrons and heat conduction of the bulk 
electrons), and CQ time $\taucq$, in accordance with previous studies 
\citep{pusztai2023, ekmark2024}. 
Both the runaway current and the transported heat load fraction should be minimized in order to avoid 
damage to the plasma facing components. 
More specifically, a transported heat load fraction, i.e. the fraction of the initial plasma kinetic 
energy which has been lost from the plasma due to energy transport, of less than \SI{10}{\percent} would 
be desirable during a disruption \citep{hollmann2015}. 
Regarding the runaway current, there is no strict upper limit, but a runaway current of less than 
\SI{150}{kA}, as previously used for ITER \citep{pusztai2023, ekmark2024}, would be ideal, while even a 
runaway current of up to \SI{1}{MA} might be tolerable. 
Note that the plasma facing components in SPARC are not actively cooled. 
The risks posed by runaway strikes are tile surface degradation and the production of impurities that may 
affect subsequent plasmas.
The runaway current has been quantified by its value at the time when it reaches \SI{95}{\percent} of the 
remaining total plasma current ($\Ire(t_{\Ire=0.95\Ip})$), unless this happens after the occurrence of the 
maximum runaway current ($\Ire=\max_t\Ire$), in which case the latter will be used, as motivated in 
appendix A.1 of \citep{ekmark2024}.

The duration of the CQ, or CQ time, is also an important metric, as too short CQ times can lead to 
mechanical stresses due to induced eddy currents in the plasma surrounding structures, while too long CQ 
times can lead to intolerably large halo currents in plasma facing components. 
In SPARC, the range of tolerable CQ times is expected to be bounded from below and above by \SI{3.2}{ms} 
\citep{sweeney2020} and \SI{40}{ms} according to the ITPA Disruption Database \citep{eidietis2015}. 
The lower bound of \SI{3.2}{ms} has been chosen because all SPARC components are designed to withstand an 
exponential CQ with characteristic decay time $\tau=\SI{1.385}{ms}$ (equivalent to a \SI{3.2}{ms} linear CQ 
following the ITPA 80-\SI{20}{\percent} convention). 
Furthermore, using the ITPA 80-\SI{20}{\percent} convention, we evaluate the CQ time through extrapolation, 
i.e. ${\taucq=(t_{\Iohm=0.2\Ipin}-t_{\Iohm=0.8\Ipin})/0.6}$ if the Ohmic current drops below 
\SI{20}{\percent} of the initial plasma current ($\Ipin=\SI{8.7}{MA}$) during the simulation and, 
otherwise, ${\taucq=(t_{\rm final}- t_{\Iohm=0.8\Ipin})/(0.8-\Iohmfin/\Ipin)}$. 
Note that we denote the CQ time $\taucq$ to signify that this CQ time corresponds to the decay of the 
Ohmic current, rather than the total plasma current. 
Additionally, in order to avoid favouring simulations with an incomplete CQ, we also want to minimize the 
final Ohmic current in the cost function. 
Furthermore, a high final Ohmic current has the potential to be converted into runaway current, and thus 
introduces an uncertainty for such disruptions. 

We use the same cost function $\Lcf$ as that devised by \cite{ekmark2024}, which was designed with the main 
aim of distinguishing between successful ($\Lcf<1$) and unsuccessful ($\Lcf>1$) disruption mitigation. 
It takes the form
\begin{subequations}
\begin{align}
    \Lcf&=\frac12\sqrt{f_\Ire^2+f_\Iohm^2 + f_\etac^2+f_\taucq^2},\\
    f_i&=\label{eq:Lcf_imp}
        \begin{cases}
            (x_i)^{k_i}, &\text{if } x_i\leq1,\\
            k_i(x_i-1)+1, &\text{if } x_i\geq1,
        \end{cases}
\end{align}
\end{subequations}
where ${x_I=I/\SI{150}{kA}}$ and ${x_\etac=\etac/\SI{10}{\percent}}$, since they should be minimized. 
For the CQ time, the goal is not to minimize it, but rather to contain it within a certain interval, 
and therefore we use ${x_\taucq=\abs{\taucq-\SI{21.6}{ms}}/\SI{18.4}{ms}}$. 
Using this form causes the optimal value of $\taucq$ to be located at the middle of the interval at 
$(3.2+40)/2=\SI{21.6}{ms}$, but in reality, it is not necessarily better to have, e.g. 
$\taucq=\SI{20}{ms}$ than $\taucq=\SI{10}{ms}$. 
For this reason we use $k_\taucq=6$, as for CQ times well within the interval of safety $f_\taucq\ll1$. 
Furthermore, we use $k_I=1$ for the currents and $k_\etac=3$ for the transported heat loss. 
Thus, using the form \eqref{eq:Lcf_imp} ensures that, when all the cost function values are within their 
acceptable limits, minimizing the currents will be favoured over minimizing the transported heat loss, 
which in turn will be favoured over having $\taucq$ close to \SI{21.6}{ms}. 

In order to avoid unnecessarily long simulations, the \DREAM{} simulations are stopped when all the cost 
function quantities can be determined and only vary negligibly with time. 
More specifically, this happens when $\Iohm<0.2\Ipin$ and either $\Irenf$ or $\Iremax$ has occurred. 
Additionally, we require that either $\Iohm<\SI{100}{kA}$ at the end of the simulation or that we have 
simulated for longer than $2\taucq$. 
The simulation will also be terminated if $\Ire<\SI{150}{kA}$ and $\Iohm<\SI{100}{kA}$, even though 
neither of $\Irenf$ and $\Iremax$ has occurred. 
Using these termination conditions means that the final Ohmic current will not play a big role in how the 
cost function varies, as for most simulations it will be $\sim\SI{100}{kA}$. 
It will only play a significant role if $\Iohm\gg\SI{150}{kA}$. 
Another consequence of this is that $\Lcf\geq0.5\times(\SI{100}{kA}) / (\SI{150}{kA})\approx0.33$, which 
does limit the interpretability of the cost function for low cost function values, but this was deemed 
less important than the efficiency gain of using such termination conditions.

The optimization parameters are the densities of the injected material, notably using logarithmic scales 
to enable studying MMI densities over several orders of magnitude. 
The optimization bounds can be found in table~\ref{tab:opt} together with the limits of safety for the 
disruption figures of merit. 
In SPARC, the current MGI design allows for injected quantities resulting in \deuterium{} densities 
${<\SI{4.8e22}{m^{-3}}}$ and \neon{} and \argon{} densities ${<\SI{4.7e21}{m^{-3}}}$, while currently there 
is no plan to use \helium{} as an MGI gas. 
These upper design limits of the densities have been evaluated under the assumption that the injected gas 
is distributed uniformly in the whole vacuum vessel, with a volume of \SI{45}{m^3}. 
In fact, the upper design limits regard injected particle numbers, not particle densities, and arise from the 
fuel processing. 
Since we only simulate the plasma, with a plasma volume of \SI{20}{m^3}, assuming uniform distribution of 
the injected material means that the nominal upper bound corresponds to \SI{44}{\percent} assimilation of 
\deuterium. 
In ASDEX Upgrade, assimilation of \SI{40}{\percent} has been reached, making this a reasonable limit to 
consider \citep{pautasso2015}. 
Thus notably, while our optimization will explore noble gas densities within their allowed ranges, we 
explore higher \deuterium{} densities than will be possible to achieve, according to the current design. 
Optimizations have been performed for disruptions during \deuterium{} operation caused by \deuterium{} and 
\neon{} injections, as well as disruptions during \deuteriumtritium{} operation caused by \deuterium{} in 
combination with \neon, \argon{} or \helium{} injections. 
For each optimization, 400 simulations have been performed. 

\begin{table}
	\centering
	\caption{Bounds for successful mitigation of the disruption figures of merit used for the cost 
    function, as well as the bounds used for the injected material densities in the optimizations. }
    \label{tab:opt}
    \begin{tabular}{l c c c c c c c }
		\toprule
        & $\Ire$ [kA] & $\etac$ [\%] & $\taucq$ [ms] & $\nD$ [m$^{-3}$] 
        & $\nHe$ [m$^{-3}$] & $\nNe$ [m$^{-3}$] & $\nAr$ [m$^{-3}$] \\
        \midrule
        Lower bound & -- & -- & 3.2 & $3.16\times10^{19}$ & $3.16\times10^{19}$ & 
        $1\times10^{18}$ & $1\times10^{17}$ \\
        Upper bound & 150 & 10 & 40 & $1\times10^{23}$ & $1\times10^{23}$ & 
        $3.16\times10^{21}$ & $3.16\times10^{20}$ \\
	\end{tabular}
\end{table}

In this work, we have used a Bayesian optimization strategy using Gaussian processes, meaning that it is 
assumed that the nature of the random variables involved is Gaussian. 
Bayesian optimization is advantageous, compared to e.g.~a grid scan, since regions of interest are well 
resolved, while less computational resources are focused on simulations corresponding to poor disruption 
mitigation. 
In Bayesian optimization, inference based on Bayesian statistics is used to evaluate a probability 
distribution function for the cost function based on a set of samples. 
Through this probability density function, an estimation for the cost function landscape can be obtained 
through the mean of the probability density $\mu$ and an error estimate can be obtained from the 
covariance. 
Furthermore, an acquisition function is needed in order to determine how to choose new samples during the 
optimization, and for this, we have used the expected improvement acquisition function, which chooses the 
sample that maximizes the expected improvement. 
Here, the Bayesian optimization was performed using the Python package by \citet{pybayesopt}. 
For the Gaussian processes, the Matérn covariance kernel with smoothness parameter $\nu=3/2$ was used. 

\section{Disruption mitigation optimization}
Here, we present the results from the disruption mitigation optimization with regard to the MMI densities. 
In \S\ref{sec:resDD}, we optimize disruption mitigation for a pure \deuterium{} plasma using MMI of 
\deuterium{} and \neon. 
We consider cases with and without generation of REs from Compton scattering due to photon flux from the 
DD neutron bombardment of the wall. 
In \S\ref{sec:resDT}, we instead focus on \deuteriumtritium{} operation, and then account for REs generated 
both from Compton scattering and tritium beta decay. 
For disruptions of \deuteriumtritium{} plasmas, aside from \neon, we also consider \helium{} and \argon{} 
MMI in combination with \deuterium. 

\subsection{Deuterium operation}\label{sec:resDD}
For pure \deuterium{} plasmas, the RE generation from Compton scattering is often neglected due to the 
low neutron energy of the neutronic DD reaction and low cross section of the DD fusion reaction 
\citep{pusztai2023, ekmark2024}.
When the RE seed is only generated by hot-tail and Dreicer generation, it is easier to suppress the 
formation of a significant runaway current, making the mitigation of such scenarios more attainable.  
In this section, we have optimized the MMI densities of \deuterium{} and \neon{} for pure \deuterium{} 
plasmas, both with and without the Compton seed.

\begin{figure}
    \centering
    \begin{subfigure}[t]{4.435cm}
        \begin{overpic}[width=4.435cm]{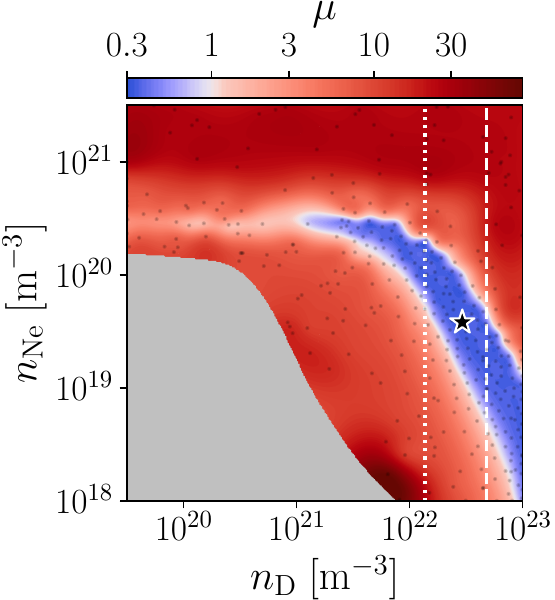} 
            \put(22, 23){(a)}
            \put(33, 26){DD}
            \put(33, 20){w/o $\gamma_{\rm C}$}
        \end{overpic}
        \phantomcaption
        \label{fig:opt_D}
    \end{subfigure}
    \begin{subfigure}[t]{4.435cm}
        \begin{overpic}[width=4.435cm]{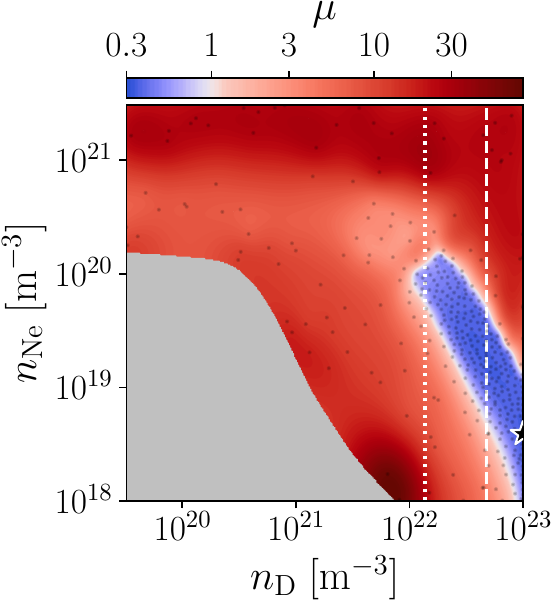}
            \put(22, 23){(b)}
            \put(33, 26){DD}
            \put(33, 20){w/ $\gamma_{\rm C}$}
        \end{overpic}
        \phantomcaption
        \label{fig:opt_D_C}
    \end{subfigure}
    \begin{subfigure}[t]{4.435cm}
        \begin{overpic}[width=4.435cm]{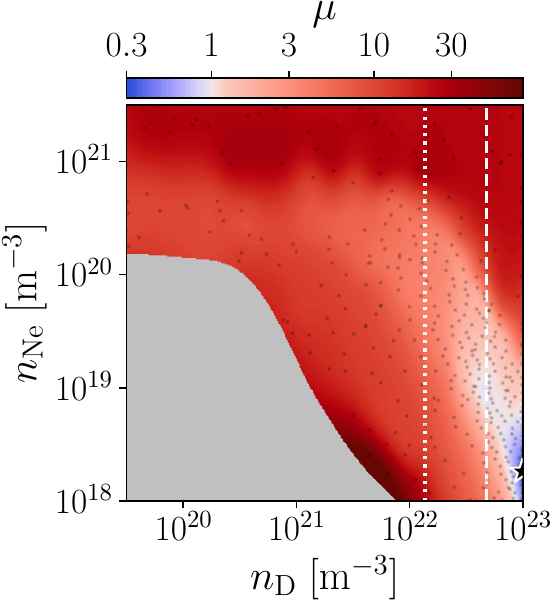}
            \put(22, 23){(c)}
            \put(33, 26){DT, w/}
            \put(33, 20){$\gamma_{\rm C}+\gamma_{\rm T}$}
        \end{overpic}
        \phantomcaption
        \label{fig:opt_DT}
    \end{subfigure}
    \caption{Logarithmic contour plots of the cost function estimate $\mu$ for \deuterium{} operation
    \mbox{(a) without Compton generation} and (b) with DD-induced Compton generation, as well as (c) for 
    \deuteriumtritium{} operation with RE generation from both DT-induced Compton scattering and tritium beta 
    decay.
    Note that the colour mapping is adapted such that blue shades represent regions of safe operation. 
    The black star indicates the optimal samples, while the black dots indicate all optimization samples, 
    and the upper design limit of the \deuterium{} density during MGI in SPARC is indicated by the dashed 
    vertical line for \SI{44}{\percent} assimilation ($\nD=\SI{4.8e22}{m^{-3}}$) and the dotted vertical 
    line for \SI{10}{\percent} assimilation.
    The grey area covers the region of incomplete TQ.}
    \label{fig:opt}
\end{figure}
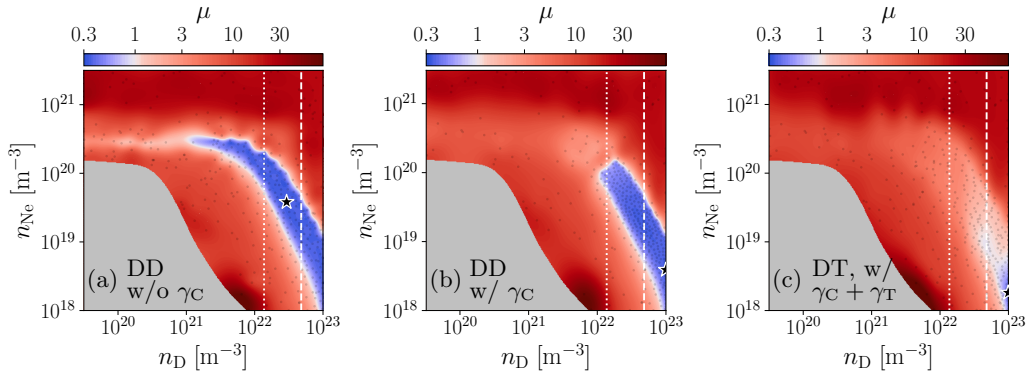

The estimated cost function landscape obtained from the optimization samples without generation from 
Compton scattering is shown in figure~\ref{fig:opt_D}.
There is a relatively large band of successful disruption mitigation scenarios, as indicated by the blue 
shaded region in the plot, which stretches from ${\nD\sim\SI{e21}{m^{-3}}}$ and 
${\nNe\sim\SI{3e20}{m^{-3}}}$ to ${\nD\sim\SI{e23}{m^{-3}}}$ and ${\nNe\sim\SI{3e18}{m^{-3}}}$.
Parts of this safe region even stretch below the upper design limit with \SI{10}{\percent} assimilation. 
As indicated by the deep blue shade of this region of safety, the majority of this region has 
$\mu\approx0.3$, which is the lower bound for the cost function due to the termination conditions we use 
for the CQ simulation, as discussed in \S\ref{sec:opt}. 
Thus, for the major part of the safe region, the final Ohmic current dominates the cost function values. 
The optimum can be found in the middle of this safe region at ${\nD\approx\SI{2.9e22}{m^{-3}}}$ and 
${\nNe\approx\SI{3.8e19}{m^{-3}}}$, 
corresponding to \SI{27}{\percent} assimilation of the upper design limit. 
At this optimal sample, no runaway current is generated, while $\taucq\approx\SI{7.8}{ms}$ and 
$\etac\approx\SI{4.5}{\percent}$, meaning that all the figures of merit we use to quantify the disruption 
are well within their limits of safe operation. 

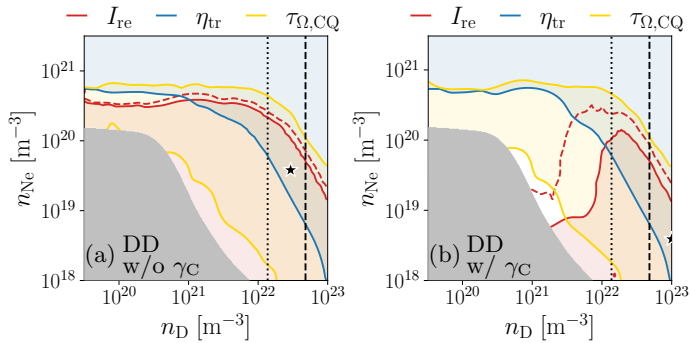
\begin{figure}
    \centering
    \begin{subfigure}[t]{4.44cm}
        \begin{overpic}[width=4.44cm]{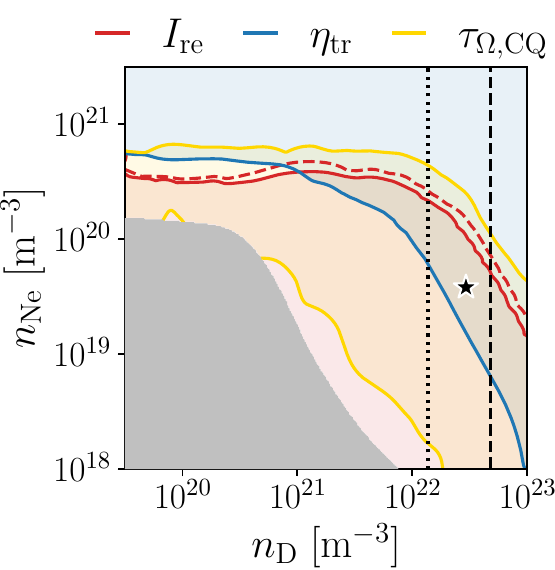} 
            \put(22, 23){(a)}
            \put(33, 26){DD}
            \put(33, 20){w/o $\gamma_{\rm C}$}
        \end{overpic}
        \phantomcaption
        \label{fig:conts_D}
    \end{subfigure}
    \begin{subfigure}[t]{4.44cm}
        \begin{overpic}[width=4.44cm]{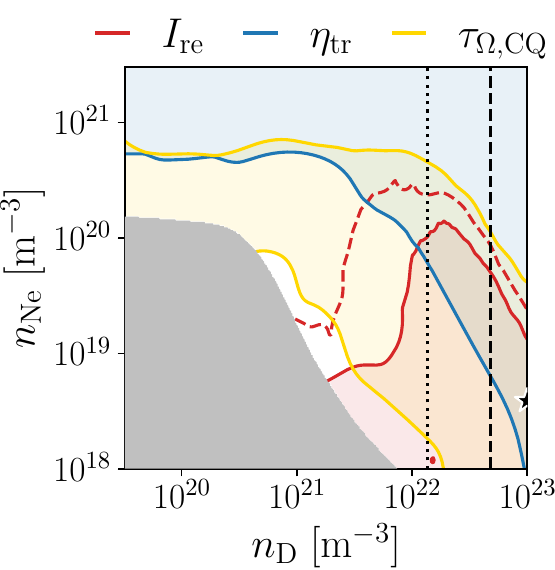} 
            \put(22, 23){(b)}
            \put(33, 26){DD}
            \put(33, 20){w/ $\gamma_{\rm C}$}
        \end{overpic}
        \phantomcaption
        \label{fig:conts_D_C}
    \end{subfigure}
    \caption{Regions of safe operation (shaded) with regards to $\Ire$ (red), $\etac$ (blue) and 
    $\taucq$ (yellow) for \deuterium{} operation (a) without Compton generation and (b) with Compton 
    generation.
    Additionally, the red dashed line indicates where the runaway current is 1 MA, bounding the tolerable 
    region of operation.
    The optimal sample is indicated by a star, the upper design limit of the \deuterium{} density during 
    MGI in SPARC is indicated by the dashed vertical line for \SI{44}{\percent} assimilation 
    ($\nD=\SI{4.8e22}{m^{-3}}$) and the dotted vertical line for \SI{10}{\percent} assimilation.
    The grey area covers the region of incomplete TQ.
    }
    \label{fig:contsNA}
\end{figure}

In figure~\ref{fig:conts_D}, the regions of safe operation for each individual figure of merit across the 
explored density space in figure~\ref{fig:opt_D} are shown, that is for the optimization of \deuterium{} 
operation without Compton generation. 
The more detailed landscapes of the runaway current, transported heat loss and CQ time are presented in 
figure~\ref{fig:comps_D} in appendix \ref{app:land}. 
The safe region for the runaway current corresponds to lower \neon{} densities, and spans approximately the 
lower half of the explored density space. 
Close to the boundary of this safe region, the gradient in density space is very high, as the contour line 
for $\Ire=\SI{150}{kA}$ (solid red) and the one for $\Ire=\SI{1}{MA}$ (dashed red) are very close to each 
other. 
Thus, increasing the \neon{} density slightly around this boundary will cause a sharp increase in the RE 
density. 
On the other hand, for high values of the MMI densities, the transported heat fraction is within its limit 
of safe operation -- the region of $\etac<\SI{10}{\percent}$ covers the upper right portion of the explored 
density space. 

When RE generation from DD neutron-induced Compton scattering is considered, the region of safe operation 
decreases at high \neon{} densities (figure~\ref{fig:opt_D_C}) as the regions of $\Ire<\SI{150}{kA}$ and 
$\Ire<\SI{1}{MA}$ shrink (figure~\ref{fig:conts_D_C}). 
More specifically, for the region in injected-material-density space corresponding to successful 
mitigation, the lower boundary for the \deuterium{} density is increased by an order of magnitude, while 
the upper boundary for the \neon{} denstity is decreased by a factor of $\sim2.5$. 
The location of the optimum found is significantly different to that of the optimum found when disregarding 
Compton generation; however, this is due to the cost function value being $\sim0.35$ along the valley of 
the blue region of figure \ref{fig:opt_D_C}. 
Thus, when accounting for Compton generation, the optimum in figure \ref{fig:opt_D} is equivalently 
successful in terms of mitigation to the optimum in figure \ref{fig:opt_D_C}. 
With RE generation from Compton scattering, there is also a larger separation between the contour lines of 
$\Ire=\SI{150}{kA}$ and $\Ire=\SI{1}{MA}$ (red solid and dashed in figure~\ref{fig:conts_D_C}, 
respectively). 
Thus, even during \deuterium{} operation, Compton scattering can have a significant impact on the disruption 
dynamics. 
Furthermore, this informs us that Compton scattering will play an important role during \deuteriumtritium{} 
operation, as the photon flux will be significantly larger then (see table~\ref{tab:compton}).



\subsection{Deuterium-tritium operation}\label{sec:resDT}
Disruptions are expected to be more difficult to mitigate during \deuteriumtritium{} operation, due to the generation of REs 
from DT-induced Compton scattering, which is expected to be more severe than DD-induced generation due 
to the neutrons having higher energies, and the additional generation from tritium beta decay. 
In a similar study for disruption mitigation in ITER during activated operation, it was found that there 
were no regions of safe operation in the \deuterium{} and \neon{} MMI density space explored 
\citep{ekmark2024}. 
In this section, we perform the corresponding study for SPARC, and we additionally consider \deuterium{} 
combined with \helium{} and \argon{} MMI.

\begin{table}
	\centering
	\caption{Disruption figures of merit for the simulations corresponding to the samples indicated in 
    figure \ref{fig:contsact_Ne} and \ref{fig:contsact_Ar}, both for \neon\ and \argon\ MMI.}
    \label{tab:cases}
	\begin{tabular}{c c c l c c c c c c}
		\toprule
		Marker & $\nD$ [m$^{-3}$] & $n_Z$  [m$^{-3}$] & Noble gas & $\Lcf$ & $\Ire$ [kA] 
        & $\taucq$ [ms] & $\etac$ [\si{\percent}] & $\tau_{\rm TQ}$ [ms] \\
		\midrule
        \multirow{2}{0.3cm}{\includegraphics[width=0.3cm, trim={0.6cm 0 0.6cm 0},clip]{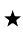}} 
        & \multirow{2}{1.4cm}{$9.9\times 10^{22}$} & \multirow{2}{1.4cm}{$1.8\times 10^{18}$} 
            &  \neon   &  $0.39$  &  $28$   &  $9.4$  &  $7.0$  &  $0.1$   \\
        & & &  \argon  &  $2.2$   &  $660$  &  $5.0$  &  $3.3$  &  $0.09$  \\
		\midrule
        \multirow{2}{0.3cm}{\includegraphics[width=0.3cm, trim={0.6cm 0 0.6cm 0},clip]{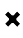}} 
        & \multirow{2}{1.4cm}{$9.9\times 10^{22}$} & \multirow{2}{1.4cm}{$1.8\times 10^{17}$} 
            &  \neon   &  $1.2$   &  $0.0$   &  $18$   &  $14$   &  $0.3$  \\
        & & &  \argon  &  $0.69$  &  $130$   &  $9.2$  &  $9.6$  &  $0.2$  \\
		\midrule
        \multirow{2}{0.3cm}{\includegraphics[width=0.3cm, trim={0.6cm 0 0.6cm 0},clip]{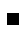}} 
        & \multirow{2}{1.4cm}{$4.0\times 10^{22}$} & \multirow{2}{1.4cm}{$1.8\times 10^{19}$} 
            &  \neon   &  $1.2$  &  $340$  &  $8.3$  &  $5.8$  &  $0.2$   \\
        & & &  \argon  &  $2.4$  &  $700$  &  $4.3$  &  $2.1$  &  $0.1$  \\
		\midrule
        \multirow{2}{0.3cm}{\includegraphics[width=0.3cm, trim={0.6cm 0 0.6cm 0},clip]{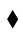}} 
        & \multirow{2}{1.4cm}{$1.0\times 10^{22}$} & \multirow{2}{1.4cm}{$1.8\times 10^{20}$} 
            &  \neon   &  $4.2$  &  $1300$  &  $6.2$  &  $4.7$  &  $0.3$   \\
        & & &  \argon  &  $21$   &  $6400$  &  $0.38$ &  $3.6$  &  $0.04$  \\
		\midrule
        \multirow{2}{0.3cm}{\includegraphics[width=0.3cm, trim={0.6cm 0 0.6cm 0},clip]{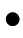}} 
        & \multirow{2}{1.4cm}{$3.2\times 10^{22}$} & \multirow{2}{1.4cm}{$5.6\times 10^{18}$} 
            &  \neon   &  $2.5$  &  $280$   &  $12$   &  $22$   &  $0.4$  \\
        & & &  \argon  &  $1.9$  &  $570$   &  $6.0$  &  $6.6$  &  $0.3$  \\
	\end{tabular}
\end{table}

\begin{figure}
    \centering
    \begin{subfigure}[t]{4.435cm}
        \begin{overpic}[width=4.435cm]{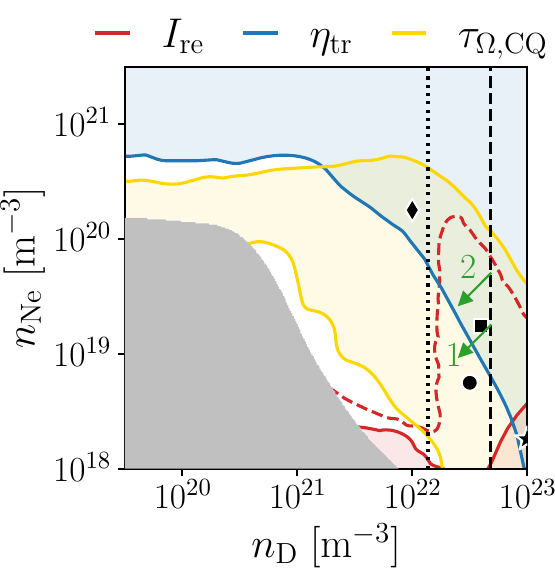}
            \put(22, 23){(a)}
            \put(33, 26){DT,}
            \put(33, 20){Ne MMI}
        \end{overpic}
        \phantomcaption
        \label{fig:contsact_Ne}
    \end{subfigure}
    \begin{subfigure}[t]{4.435cm}
        \begin{overpic}[width=4.435cm]{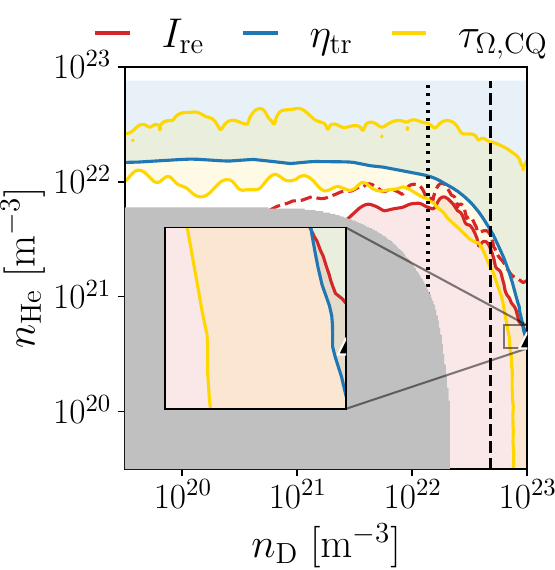}
            \put(22, 20){{(b) DT, He MMI}}
        \end{overpic}
        \phantomcaption
        \label{fig:contsact_He}
    \end{subfigure}
    \begin{subfigure}[t]{4.435cm}
        \begin{overpic}[width=4.435cm]{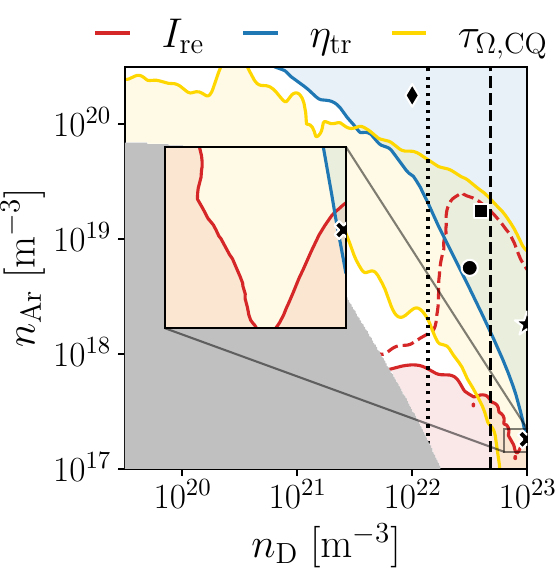} 
            \put(22, 20){(c) DT, Ar MMI}
        \end{overpic}
        \phantomcaption
        \label{fig:contsact_Ar}
    \end{subfigure}
    \caption{Regions of safe operation (shaded) with regards to $\Ire$ (red), $\etac$ (blue) and 
    $\taucq$ (yellow) for \deuteriumtritium{} operation with MMI of (a) \neon,  (b) \helium{} and (c) 
    \argon.
    Additionally, the red dashed line indicates where the runaway current is 1 MA, bounding the tolerable 
    region of operation.
    The markers indicate the cases in table~\ref{tab:cases}, while the optimal sample is indicated by a 
    star in (a), a triangle in (b) and a cross in (c).
    The upper design limit of the \deuterium{} density during MGI in SPARC is indicated by the dashed 
    vertical line for \SI{44}{\percent} assimilation ($\nD=\SI{4.8e22}{m^{-3}}$) and the dotted vertical 
    line for \SI{10}{\percent} assimilation. 
    If the assimilation is lower than \SI{44}{\percent}, the expected outcome of experiments using MMI 
    densities along the line indicating \SI{44}{\percent} would be shifted in the direction of the green 
    arrows. 
    The grey area covers the region of incomplete TQ. 
    }
    \label{fig:contsact}
\end{figure}

In figure~\ref{fig:opt_DT}, the estimated landscape of the cost function is plotted on a logarithmic scale 
for MMI of \deuterium{} and \neon. 
Notably, there is a region of safe operation for ${\nD\sim\SI{e23}{m^{-3}}}$ and 
${\nNe\sim\SI{2e18}{m^{-3}}}$, but no region of safe operation can be found below the upper design limit 
for the \deuterium{} density for SPARC at \SI{44}{\percent} assimilation, namely $\nD=\SI{4.8e22}{m^{-3}}$. 
Thus, figure~\ref{fig:opt} shows how the region of safety shrinks as the activated generation 
mechanisms play a more significant role -- first showing for \deuterium{} operation without activated 
sources in \ref{fig:opt_D}, second for when DD-induced Compton generation 
($\Gamma_{\rm flux}\sim \SI{e15}{\per\meter\squared\per\second}$) is included in \ref{fig:opt_D_C}, and 
finally for \deuteriumtritium{} operation with generation from both tritium beta decay and DT-induced 
Compton scattering ($\Gamma_{\rm flux}\sim \SI{e18}{\per\meter\squared\per\second}$) in \ref{fig:opt_DT}. 
This region of safe operation corresponds to the overlap between the regions of safety for the runaway 
current, which favours low \neon{} densities, and the transported heat loss, which favours high \neon{} 
densities, as shown in figure~\ref{fig:contsact_Ne}. 
The optimal sample is located at ${\nD\approx\SI{1e23}{m^{-3}}}$ and ${\nNe\approx\SI{2e18}{m^{-3}}}$, 
indicated by a star in figures  \ref{fig:opt_DT} and \ref{fig:contsact_Ne}, and the corresponding 
disruption simulation produced a runaway current of \SI{30}{kA} and transported heat fraction of 
\SI{7}{\percent}, as noted in table~\ref{tab:cases}. 

If we consider a relaxed tolerance limit of \SI{1}{MA} for the runaway current, instead of \SI{150}{kA}, we 
find that then there exists a region of tolerable disruptions, which is similar to the region of successful 
mitigation found during the optimizations of \deuterium{} operation, both with and without the Compton 
source. 
As the Compton scattering plays a larger role in the RE dynamics, the region of safe operation shrinks, and 
this is mainly due to the reduced region of tolerable RE currents. 
This trend is clear when comparing figures \ref{fig:conts_D} (\deuterium{} without Compton generation), 
\ref{fig:conts_D_C} (\deuterium{} with DD-induced Compton generation) and \ref{fig:contsact_Ne} 
(\deuteriumtritium{}{} with DT-induced Compton scattering). 
It is probable that for \deuteriumtritium{} without any RE generation from Compton scattering, the 
corresponding optimization landscape would look similar to that of a pure \deuterium{} plasma, illustrated 
in figure \ref{fig:opt_D}. 
Simulations of the samples of table \ref{tab:cases} for a \deuteriumtritium{} plasma (with MMI of \neon{}) 
without RE generation from Compton scattering further support this claim. 
Thus, experiments during \deuterium{} operation could help inform mitigation strategies during 
\deuteriumtritium{} operation as well. 
Furthermore, using \SI{1}{MA} as an upper safety limit for the cost function instead, the optimal sample 
would be located at ${\nD\approx\SI{4e22}{m^{-3}}}$ and ${\nNe\approx\SI{2e19}{m^{-3}}}$, indicated by a 
square in figure~\ref{fig:contsact_Ne}, which is below the upper design limit of the \deuterium{} density 
at \SI{44}{\percent} assimilation and thus would be easier to achieve under the MGI design constraints. 
For these \deuterium{} and \neon{} densities, we get a runaway current of \SI{300}{kA}, which is still 
significantly smaller than \SI{1}{MA}, and transported heat fraction of \SI{6}{\percent}. 

One important aspect to consider for this optimization is how the uncertainty with regards to assimilation 
will affect the disruption mitigation. 
As assimilation is not known with certainty, it may be beneficial to choose injected quantities that may 
nominally perform somewhat poorer than the optimum, but are more robust with respect to such uncertainties.
Currently, we have assumed a uniform distribution of the injected material, corresponding to 
\SI{44}{\percent} assimilation. 
If the assimilation is $<\SI{44}{\percent}$, the densities of the injected material inside the plasma are 
lower, corresponding to ($\nD$, $\nNe$) points being shifted in the direction indicated by the 
arrows in figure~\ref{fig:contsact_Ne}. 
Choosing \neon{} density to match what is predicted to be optimal (assuming $\Ire<\SI{1}{MA}$ is 
acceptable), it is possible that the actual point in ($\nD$, $\nNe$) space would be shifted past the 
boundary of $\etac=\SI{10}{\percent}$, as indicated by the arrow labelled 1 in the figure, resulting in 
inadequate heat load mitigation. 
Figure \ref{fig:contsact_Ne} suggest that starting at a higher \neon{} density than what is predicted to 
optimal, sufficient mitigation of both the RE current and the transported fraction of the heat losses would 
be possible, even if the assimilation is less than \SI{44}{\percent}, as indicated by the arrow labelled 2 
in the figure. 

\begin{figure}
    \centering
    \begin{subfigure}[t]{13.44cm}
        \begin{overpic}[width=13.44cm]{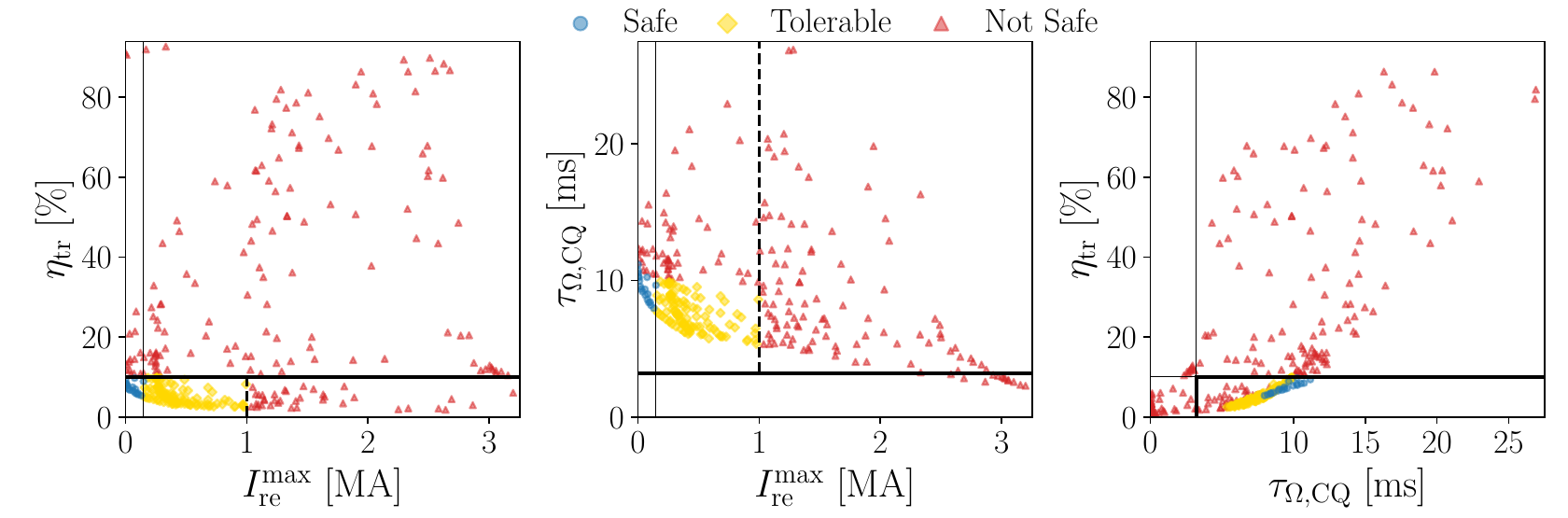}
            \put(30, 28){(a)}
            \put(62.5, 28){(b)}
            \put(95.5, 28){(c)}
        \end{overpic}
        \phantomcaption
        \label{fig:cloudsactivated_I_eta}
    \end{subfigure}
    \begin{subfigure}[t]{0cm}
        \phantomcaption
        \label{fig:cloudsactivated_I_tau}
    \end{subfigure}
    \begin{subfigure}[t]{0cm}
        \phantomcaption
        \label{fig:cloudsactivated_tau_eta}
    \end{subfigure}
    \caption{Projections of the simulation dataset to all the two-dimensional subspaces of the figure of 
    merit space $(\Ire, \etac, \taucq)$ from the \neon{} MMI optimization. 
    The intervals of safe operation for each cost function component (see table~\ref{tab:opt} for the 
    values) are indicated by the solid black lines, while the dashed black line indicates the potentially 
    tolerable upper bound of the runaway current at \SI{1}{MA}. 
    Safe simulation samples are plotted in blue, tolerable samples (namely simulations with 
    $\Ire<\SI{1}{MA}$, and otherwise safe) are plotted in yellow and unsafe samples are plotted in red. 
    This figure illustrates the trade-off between the different cost function components. }
    \label{fig:cloudsactivated}
\end{figure}

How the runaway current, transported heat loss and CQ time relate to one another for each optimization 
sample is illustrated in figure~\ref{fig:cloudsactivated}. 
The most interesting feature of this plot is that it shows the trade-off between each pair out of the three 
figures of merit.
Figure~\ref{fig:cloudsactivated_I_eta} shows that for runaway currents lower than \SI{150}{kA}, the 
transported heat loss will be larger than \SI{5}{\percent}. 
The trade-off between having a small runaway current and transported heat fraction is a consequence of two 
attributes of the injected \neon. 
The purpose of injecting \neon, or another noble gas, for mitigating a disruption is to increase the heat 
being radiated from the plasma, and thus reduce the amount of heat that is being transported to the wall. 
However, since \neon{} has $Z=10$, this also means that more electrons are also added to the plasma, which will 
increase runaway current through enhanced avalanching, as $\Gamma_{\rm ava}\propto n_{\rm e, tot}$ (where 
$n_{\rm e, tot}$ is the total electron density in the plasma). 
Thus, moving to the right along the Pareto front\footnote{This \emph{Pareto front} is the curve of optimal 
solutions with respect to the objectives $\Ire$ and $\etac$; changing their relative importance (i.e. their 
weighting) in the cost function corresponds to the optimum sweeping along this curve.} 
of the cloud  in \ref{fig:cloudsactivated_I_eta} corresponds to increasing the \neon{} density -- the 
runaway current increases while the transported heat loss decreases. 

The trade-off between the CQ time and the two other figures of merit is better, as most samples are within 
the safe interval for the CQ time. 
Notably, the safe (blue) samples are clustered around \SI{10}{ms}, which is a favourable value that 
balances the concerns of eddy current and halo current forces. 
There are some samples with $\taucq<\SI{3.2}{ms}$, which is lower than the SPARC components have been 
designed to withstand. 
In these \DREAM{} simulations, the full plasma current is almost completely converted into runaway current 
in less than \SI{3.2}{ms}, and the total plasma current does therefore not decay significantly. 
Aside from the large runaway current, as indicated by figure~\ref{fig:cloudsactivated_I_tau}, these samples 
have low values of the transported heat loss, as indicated by figure~\ref{fig:cloudsactivated_tau_eta}, 
signifying that they correspond to high \neon{} densities. 

Compared to ITER, there is thus a much more favourable compromise between the runaway current and 
transported heat fraction. 
In the ITER study, it was found that, for activated disruption simulations with a runaway current lower 
than \SI{4}{MA}, the transported heat loss would be larger than \SI{75}{\percent}, and vice versa.
Considering a number of samples from the SPARC optimization with ${\nD>\SI{e22}{m^{-3}}}$ and 
${\nNe<\SI{2e20}{m^{-3}}}$, namely the samples of table~\ref{tab:cases}, all of them present a more 
favourable compromise than this. 
This is also illustrated in figure~\ref{fig:cloudsactivated}, where many of the samples from the 
optimization can be found to have relatively low values of both $\Ire$ and $\etac$, while also having 
$\taucq$ within its acceptable range. 

The main explanation for the better trade-off in SPARC, compared to ITER, is the lower initial plasma 
current. 
In the SPARC primary reference discharge, the initial plasma current is \SI{8.7}{MA}, or \SI{60}{\percent} 
of the \SI{15}{MA} current in the ITER H-mode scenario. 
More specifically, the avalanche multiplication is exponentially sensitive to the initial plasma current, 
and thus the difference in initial plasma current is the reason the runaway current can be more 
successfully mitigated at higher \neon{} densities in SPARC, allowing for a more acceptable level of heat 
transport. 
If SPARC would have a \SI{15}{MA} initial plasma current instead of \SI{8.7}{MA}, the runaway current for 
the optimal scenario found during the optimization is \SI{2.5}{MA} instead, if we naively assume that 
the magnetic geometry is kept unchanged.

For other noble gases, the trade-off between minimizing runaway current and transported heat loss will 
still be present, but how it manifests can be different. 
We have done the same disruption optimizations for \helium{} and \argon{} MMI as well, to see how the 
disruption mitigation evolves with increasing atomic number of the noble gas. 
Notably, there is only a sliver of overlap between the safe region of the runaway current and the 
transported heat fraction for both the \helium{} and the \argon{} MMI density space, as illustrated in 
figures \ref{fig:contsact_He} and \ref{fig:contsact_Ar}. 
Using \neon{} should thus be more advantageous with regards to disruption mitigation compared to \helium{} 
or \argon, due to the much larger region of overlap shown in \ref{fig:contsact_Ne}.

More detailed plots of the landscapes of the runaway current, transported heat loss and CQ time for each 
noble gas are presented in figure~\ref{fig:comps} in appendix \ref{app:land}, where the figure of merit 
dataset projections can also be found (figure~\ref{fig:cloudsactivated_app}). 
With \helium, there is not sufficient radiative heat loss, causing the region of safety for the transported 
heat fraction to appear at very high \helium{} densities $\nHe\sim\SI{e22}{m^{-3}}$, where high runaway 
current can no longer be mitigated.
For \argon{} however, the acceptable level of heat transport is achieved even at low \argon{} densities if 
the \deuterium{} density is high enough, but it is still enough to cause significant RE avalanching. 
There is, however, a fairly large region of overlap between the region of safety for the transported heat 
fraction and the region of tolerable runaway current ($\Ire<\SI{1}{MA}$) for the \argon{} MMI, but not for 
the \helium{} MMI. 

\begin{figure}
    \centering
    \begin{subfigure}[t]{6.7cm}
        \begin{overpic}[width=6.7cm]{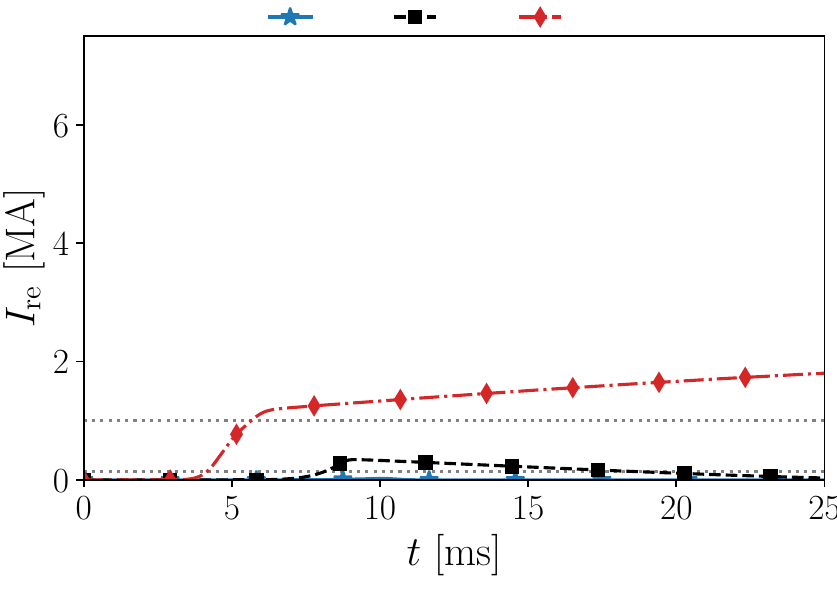} 
            \put(10, 63){(a) Ne MMI}
            \put(39, 68){
            \includegraphics[width=0.2cm, trim={0.6cm 0.1cm 0.6cm 0.1cm},clip]{figs/Markers/opt_Ne.pdf}}
            \put(53.9, 68){
            \includegraphics[width=0.2cm, trim={0.6cm 0.1cm 0.6cm 0.1cm},clip]{figs/Markers/c1.pdf}}
            \put(68.8, 68){
            \includegraphics[width=0.2cm, trim={0.6cm 0.1cm 0.6cm 0.1cm},clip]{figs/Markers/c2.pdf}}
        \end{overpic}
        \phantomcaption
        \label{fig:currevol_Ne}
    \end{subfigure}
    \begin{subfigure}[t]{6.7cm}
        \begin{overpic}[width=6.7cm]{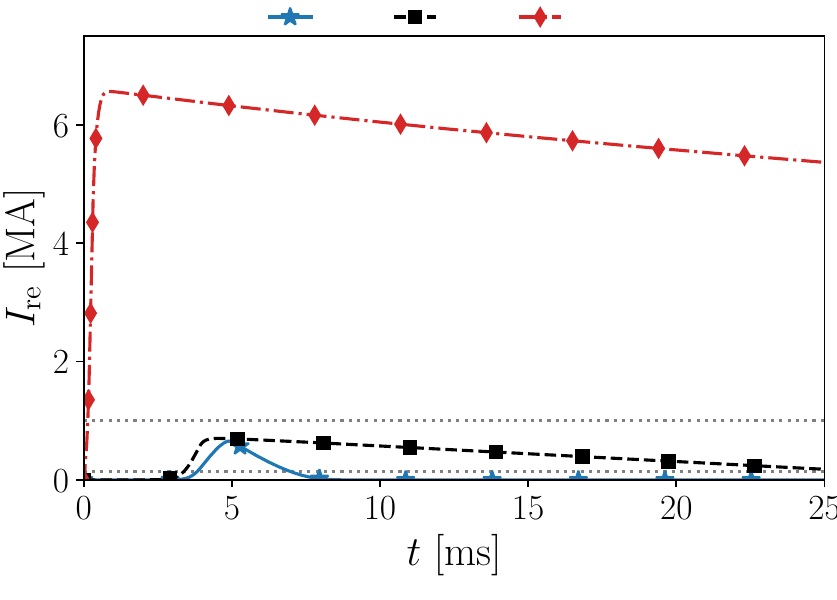} 
            \put(10, 63){(b) Ar MMI}
            \put(39, 68){
            \includegraphics[width=0.2cm, trim={0.6cm 0.1cm 0.6cm 0.1cm},clip]{figs/Markers/opt_Ne.pdf}}
            \put(53.9, 68){
            \includegraphics[width=0.2cm, trim={0.6cm 0.1cm 0.6cm 0.1cm},clip]{figs/Markers/c1.pdf}}
            \put(68.8, 68){
            \includegraphics[width=0.2cm, trim={0.6cm 0.1cm 0.6cm 0.1cm},clip]{figs/Markers/c2.pdf}}
        \end{overpic}
        \phantomcaption
        \label{fig:currevol_Ar}
    \end{subfigure}
    \caption{Current evolutions for the first (star, solid blue), third (square, dashed black) and 
    fourth sample (diamond, dash-dotted red) of table~\ref{tab:cases} for (a) \neon{} and (b) \argon{} MMI. 
    The grey dotted lines indicate $\Ire=\SI{150}{kA}$ and $\Ire=\SI{1}{MA}$.}
    \label{fig:currevol}
\end{figure}

That runaway currents are larger and transported heat fractions lower for \argon{} compared to \neon{} at 
the same atomic densities is exemplified in table~\ref{tab:cases}. 
For the transported heat fraction the difference is generally a factor $\sim2$-$3$, while for the runaway 
currents, the difference can range from a factor of $\sim2$ to orders of magnitude. 
The runaway current evolution for sample one (star), three (square) and four (diamond) from the table is 
plotted in figure~\ref{fig:currevol}.
For sample one and three, the runaway current evolution is similar for \neon{} and \argon{} MMI, even though 
the magnitudes are different. 
In both cases and for both noble gases, the runaway current peaks and then starts to decline.

\begin{figure}
    \centering
    \begin{subfigure}[t]{6.7cm}
        \begin{overpic}[width=6.7cm]{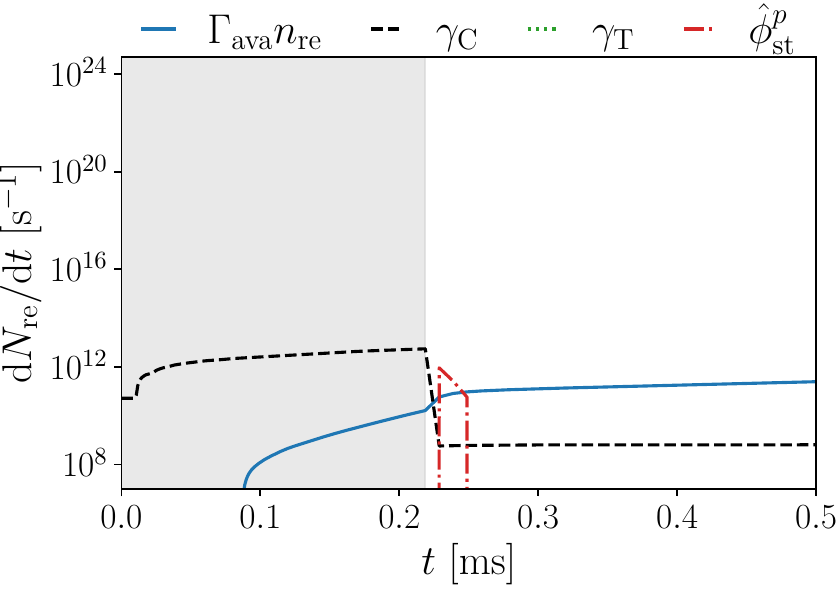} 
            \put(15, 60.5){(a) Ne MMI}
        \end{overpic}
        \phantomcaption
        \label{fig:genevol_Ne}
    \end{subfigure}
    \begin{subfigure}[t]{6.7cm}
        \begin{overpic}[width=6.7cm]{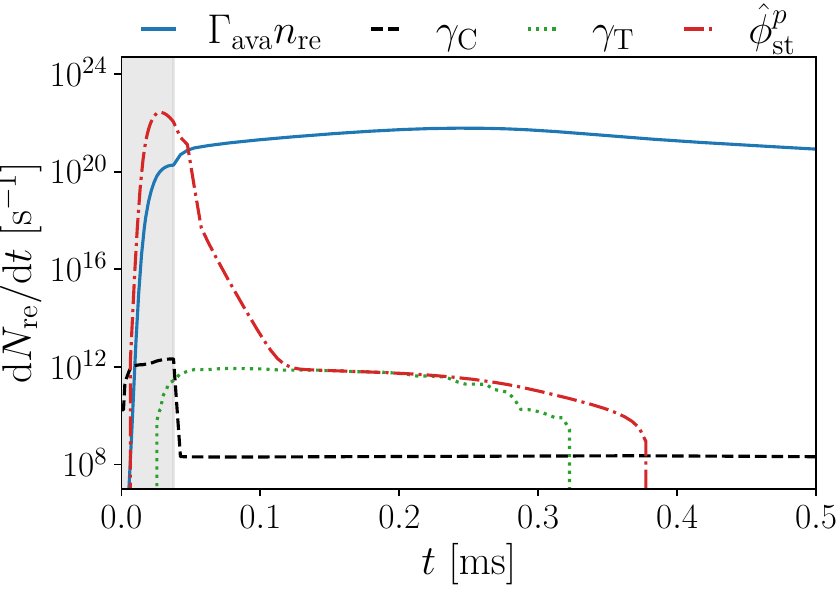} 
            \put(15, 60.5){(b) Ar MMI}
        \end{overpic}
        \phantomcaption
        \label{fig:genevol_Ar}
    \end{subfigure}
    \caption{Volume integrated generation rates for (a) third sample (square) of table~\ref{tab:cases} 
    using \neon{} as MMI material, and (b) fourth sample (diamond) of table~\ref{tab:cases} using \argon{} 
    as MMI material.
    The avalanche generation (denoted $\Gamma_{\rm ava}\nre$, note however that it is volume integrated) is 
    plotted in solid blue, generation from Compton scattering 
    (${\gamma_{\rm C}=\int_{p>p_{\rm c}}S_{\rm C}\ \dd^3\boldsymbol{p}}$) in dashed black, generation from 
    tritium beta decay (${\gamma_{\rm T}=\int_{p>p_{\rm c}}S_{\rm T}\ \dd^3\boldsymbol{p}}$) in dotted 
    green and the generation from flux across $p_{\rm c}$ 
    (${\hat{\phi}_{\rm st}^p
    =\phi_{\rm st}^p-\int_{p_{\rm c}}^{p_{\rm re}}S_{\rm C}+S_{\rm T}\ \dd^3\boldsymbol{p}}$) 
    in dash-dotted red.
    The TQ phase is indicated by the grey shaded area.
    }
    \label{fig:genevol}
\end{figure}

The runaway current dynamics are different for the fourth sample however. 
With MMI of \neon, the RE generation is slower, and it exceeds \SI{150}{kA} after \SI{3.4}{ms}, while with 
MMI of \argon{} this happens already after \SI{0.037}{ms}.  
For this \argon{} scenario, the RE generation is initially dominated by generation from momentum space 
flux, i.e. hot-tail and Dreicer generation, as illustrated in figure~\ref{fig:genevol_Ar}, but there is 
some generation from Compton scattering and tritium beta decay. 
After the TQ, the avalanche generation is dominant. 
Thus, for high amounts of injected, radiating impurities, there is significant RE seed formation from 
hot-tail and Dreicer generation, due to a fast temperature decay. 
In such a scenario, the REMC might not be effective in mitigating the RE current. 
The purpose of the REMC is to expel the RE seed before significant avalanching can occur, or to at least 
have a RE loss rate that exceeds the growth rate. 
During a scenario with strong RE seed generation, it is possible that the RE primary generation rate is 
sufficiently high for a significant RE current to be produced by avalanching, despite the additional 
transport induced by the REMC. 

It is possible to avoid large seeds from hot-tail and Dreicer generation, which for example is done in 
the third sample with MMI of \neon{} of table~\ref{tab:cases}, as illustrated in 
figure~\ref{fig:genevol_Ne}. 
Here, there is no significant generation from hot-tail and Dreicer during the TQ, and no generation at all 
from tritium beta decay, indicated by the absence of a green curve in the figure, due to 
$p_{\rm c}>p_{\rm max}$ throughout the simulation, where $p_{\rm max}$ is the maximum possible momentum for 
electrons emitted in beta decay. 
The generation from Compton scattering, however, is at around the same level as in 
figure~\ref{fig:genevol_Ar}, though for a longer period of time due to the longer duration of the TQ. 
The Compton scattering thus still generates a large enough RE seed that the runaway current, through 
avalanche generation, grows to $\SI{340}{kA}$. 
Note that the short period of significant RE generation from flux through the upper boundary $p_{\rm re}$ 
of the superthermal grid that happens after the TQ is still due to Compton generation. 
It is however delayed due to the time it takes for the REs generated at the end of the TQ with 
$p_{\rm c}<p<p_{\rm re}$ to be accelerated to $p_{\rm re}$.

Importantly, the RE seed generated in the third sample of table \ref{tab:cases} is mainly produced by the 
Compton scattering during the TQ. 
Using a reduction factor of $10^3$, as was done by \citet{vallhagen2024} and \citet{ekmark2024}, or 
$10^2$, instead of $10^4$ only changes the maximum RE current by $\SI{10}{kA}$ or 
$\SI{90}{kA}$, respectively. 
Neglecting the delayed photon flux altogether had no significant effect on the generated RE current. 
More generally, using a reduction factor of $10^3$, the maximum runaway currents are at most increased by 
$\sim\SI{50}{kA}$ for the samples considered in table \ref{tab:cases}. 
Thus, for SPARC, the RE generation from delayed photon flux is not an important effect. 
However, since it is not possible to know this result \emph{a priori}, accounting for a finite delayed 
photon flux is useful; indeed the results of \citep{vallhagen2024} indicate that the Compton seed created 
after the TQ losses may put a multi-MA lower bound on the RE currents in DT plasmas of ITER.

\section{Conclusions}
We have studied disruption mitigation in the primary reference discharge of SPARC using MMI of \deuterium{} 
and noble gases. 
We do so by optimizing the MMI densities to find parameter regions with acceptably low 
levels of heat transport to the wall and runaway currents, as well as acceptable CQ durations. 
For the first time, we also consider the primary RE generation caused by Compton scattering in a tokamak 
activated by DD neutrons. 
We find that during \deuterium{} operation, disruptions can be successfully mitigated using MMI of 
\deuterium{} and \neon, even when accounting for the DD-induced Compton scattering. 
We observe, however, that this primary generation process does decrease the $\nD$-$\nNe$ parameter region 
corresponding to successful mitigation, signifying that generation from Compton scattering can play an 
important role even during \deuterium{} operation. 
Figures~\ref{fig:opt_D} and \ref{fig:opt_D_C} display a possible path in parameter space that could be 
mapped out through \deuterium{} plasma experiments, i.e.~performing several experiments with injected 
\deuterium{} and \neon{} densities chosen from the valley of the safe region of operation for a pure 
\deuterium{} plasma. 
This could give confidence in following the same trajectory in \deuteriumtritium{} plasmas as well. 

For \deuteriumtritium{} operation, we also find a region of successful mitigation with 
${\Ire<\SI{150}{kA}}$, but it requires \deuterium{} densities above the upper design limit of 
$\nD=\SI{4.8e22}{m^{-3}}$ assuming \SI{44}{\percent} assimilation. 
It is however possible to achieve tolerable disruptions, with ${\Ire<\SI{1}{MA}}$, below this upper 
design limit. 
We thus find a more favourable compromise for acceptable values of the runaway current and transported heat 
fraction in SPARC compared to ITER, as in ITER it was not possible to achieve ${\Ire<\SI{4}{MA}}$ and 
${\etac<\SI{75}{\percent}}$ simultaneously \citep{ekmark2024}. 
The main reason for the more favourable mitigation in SPARC is the lower initial plasma current, leading to 
lower avalanching. 

For \deuteriumtritium{} operation, we also studied MMI of \helium{} and \argon{} in addition to \neon, but 
out of the three, \neon{} demonstrated the best trade-off between runaway current and transported heat. 
Out of \helium{} and \argon, \argon{} resulted in better mitigation, as the region for tolerable runaway 
currents ($\Ire<\SI{1}{MA}$) was significantly larger for \argon{} than for \helium. 
However, the region of tolerable runaway currents was larger still for \neon, compared to \argon, 
and there was a larger region of overlap between safe values of the runaway current ($\Ire<\SI{150}{kA}$) 
and the transported heat loss ($\etac<\SI{10}{\percent}$). 
These results specifically regard single material injection for RE avoidance. 
According to recent work by \cite{hollmann2023}, for a double injection scheme, using \argon{} instead of 
\neon{} during the initial injection, could be more beneficial for benign termination. 
However, recent work by \citet{sheikh2024} and \citet{hoppe2025} suggests that above a vessel pressure of 
\SI{1}{Pa}, benign termination may no longer be effective, corresponding to a relatively low post-injection 
plasma \deuterium{} density of \SI{6e20}{m^{-3}} in SPARC, which would greatly constrain the parameter 
space studied herein. 
Further the simulation work by \citeauthor{hollmann2023} suggests that the high current density in SPARC 
also challenges benign termination. 
As such optimizing the primary injection together with the REMC might be preferable. 

One point of note for the disruption simulations in this work is the relatively simple MMI model used. 
In the model, the material is instantly and uniformly distributed in the plasma, and the transport due to 
magnetic field line stochastization is activated simultaneously. 
In reality, the material would require time to penetrate the plasma from the edge, which 
could significantly impact the plasma evolution, and runaway electron dynamics. 
Accounting for a more realistic model of the injected material deposition, with temporal and spatial 
variations, would mean that plasma cooling would happen earlier farther out from the magnetic axis.
This could trigger magnetic perturbations in the core before the material has penetrated that far into the 
plasma. 
Notably, this would impact runaway dynamics through transport, the hot-tail generation -- as the cooling 
dynamics could be significantly different --, as well as the other seed generation mechanisms -- as the 
evolution of the critical and Dreicer fields could be affected. 
More specifically, the hot-tail generation could be overestimated by our model, due to the drastic 
temperature decay in the core caused by instantaneous deposition. 
On the other hand, the mitigating effects of the injected material could also be overestimated for the same 
reason. 
Assessing these effects would require self-consistent modelling of the material penetration, MHD 
instabilities and runaway dynamics, and is outside the scope of the paper. 
It is however worth pointing out that the pre-TQ duration is estimated to be $\sim\SI{2}{ms}$ according to 
the ITPA Disruption Database \citep{eidietis2015}, and  the disruption mitigation system currently under 
construction for SPARC should be able to deliver a \deuterium{} density of \SI{4.4e22}{m^{-3}} during this 
time. 
Thus, for pure \deuterium{} operation, the densities required for successful mitigation are attainable 
during the pre-TQ with the current system. 
For \deuteriumtritium{} plasmas, the densities required could be reached, but the fuel processing systems 
may constrain the maximum allowable injection. 

In this study, we have not accounted for other relevant effects that could help mitigate the disruptions, 
such as plasma scrape-off \citep{wang2025,vallhagen2025}, secondary material injection into a developed RE 
beam \citep{carlos2019,reux2021} and the use of a REMC \citep{tinguely2021}. 
Previous studies suggest that these effects can drastically limit the impact or size of the runaway 
current. 
This means that the runaway currents found in this study represent a worst-case scenario for what could 
happen during a disruption in SPARC.
Other effects which we have neglected which could change the figure of merit landscape are the impurity 
deposition profile and its time evolution, impurity injection duration, equilibrium evolution and RE 
ionization.

In summary, we find that acceptable levels of runaway current and transported heat losses can be achieved 
in \deuterium{} operation, even without the REMC. 
This applies also to \deuteriumtritium{} plasmas if the injected \deuterium{} densities can be increased 
above the nominal upper limit for the SPARC MGI system or assimilation closer to \SI{70}{\percent} 
(corresponding to $\nD=\SI{e23}{m^{-3}}$) can be achieved. 
Considering how massive material injection can expand the safe mitigation space when used in combination 
with the REMC, and how it might be used to relax requirements on the mitigation coil 
system, appear as  interesting avenues for future investigation. 

\section*{Acknowledgements} 
The authors are grateful to O Vallhagen, E Nardon and S Newton for fruitful discussions and to H Boyd for
providing the MCNP data. 

\section*{Funding} 
This work was supported by the Swedish Research Council (Dnr.~2022-02862 and 2021-03943), and by the Knut 
and Alice Wallenberg foundation (Dnr.~2022.0087 and 2023.0249).
The work has been carried out within the framework of the EUROfusion Consortium, funded by the European 
Union via the Euratom Research and Training Programme (Grant Agreement No 101052200 — EUROfusion). 
Views and opinions expressed are however those of the authors only and do not necessarily reflect those of 
the European Union or the European Commission. Neither the European Union nor the European Commission can 
be held responsible for them. 
Authors R.A. Tinguely and R. Sweeney are supported by Commonwealth Fusion Systems.

\section*{Declaration of Interests}
The authors report no conflict of interest.

\appendix

\section{Figure of merit landscapes and trade-off}\label{app:land}
\begin{figure}
    \centering
    \begin{subfigure}[t]{13.44cm}
        \begin{overpic}[width=13.44cm]{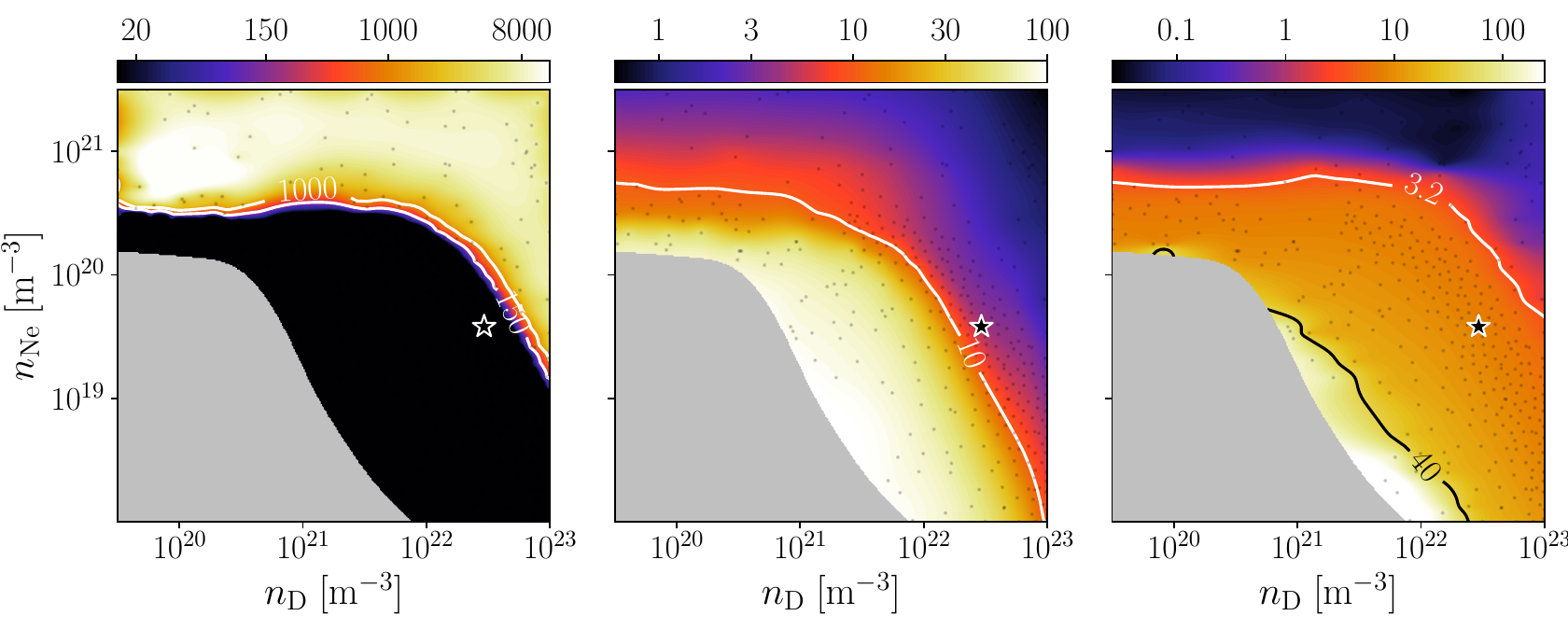}
            \put(3.5, 35){(a)}
            \put(9, 9){$\Ire$ [kA]}
            \put(40.5, 9){$\etac$ [\%]}
            \put(72, 9){$\taucq$ [ms]}
        \end{overpic}
        \phantomcaption
        \label{fig:comps_D}
    \end{subfigure}
    \begin{subfigure}[t]{13.44cm}
        \begin{overpic}[width=13.44cm]{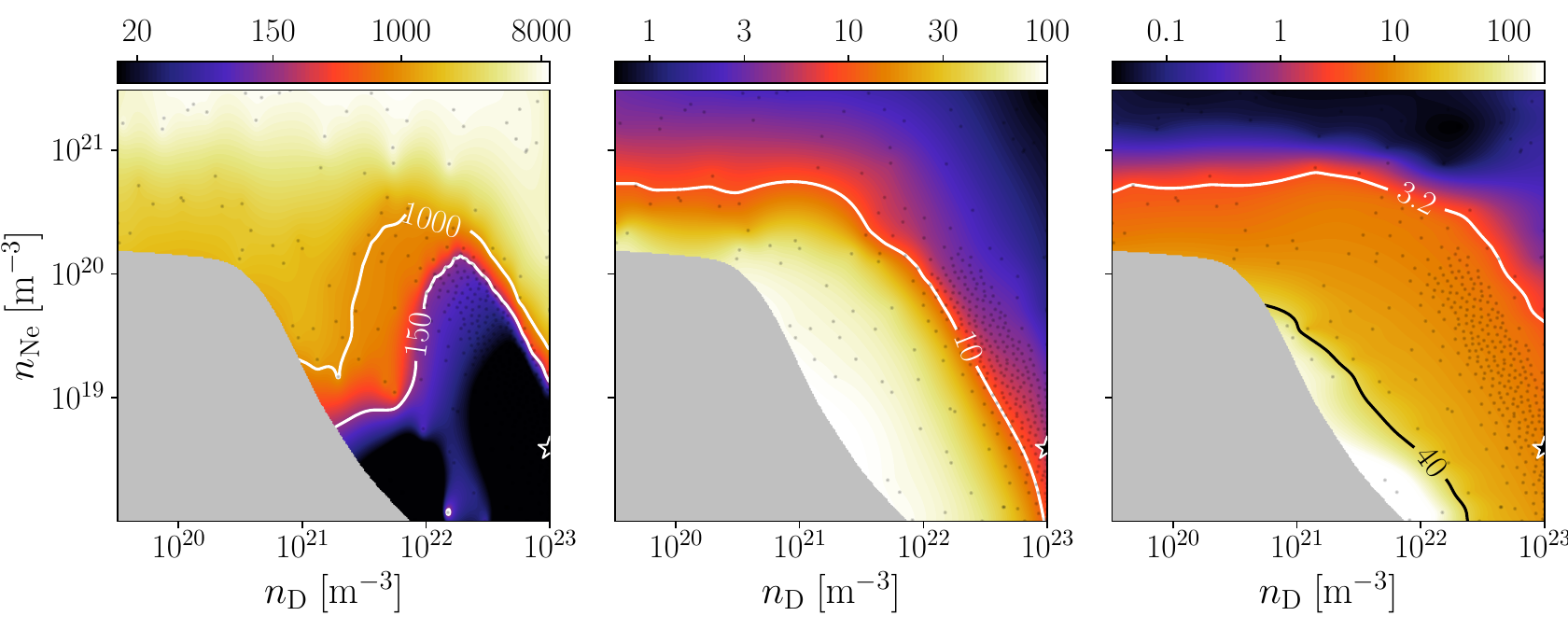}
            \put(3.5, 35){(b)}
            \put(9, 9){$\Ire$ [kA]}
            \put(40.5, 9){$\etac$ [\%]}
            \put(72, 9){$\taucq$ [ms]}
        \end{overpic} 
        \phantomcaption
        \label{fig:comps_D_C}
    \end{subfigure}
    \caption{Logarithmic contour plots of the figure of merit estimates from the optimizations of 
    \deuterium{} operation (a) without and (b) with RE generation from DD-induced Compton scattering.
    }
    \label{fig:compsNA}
\end{figure}

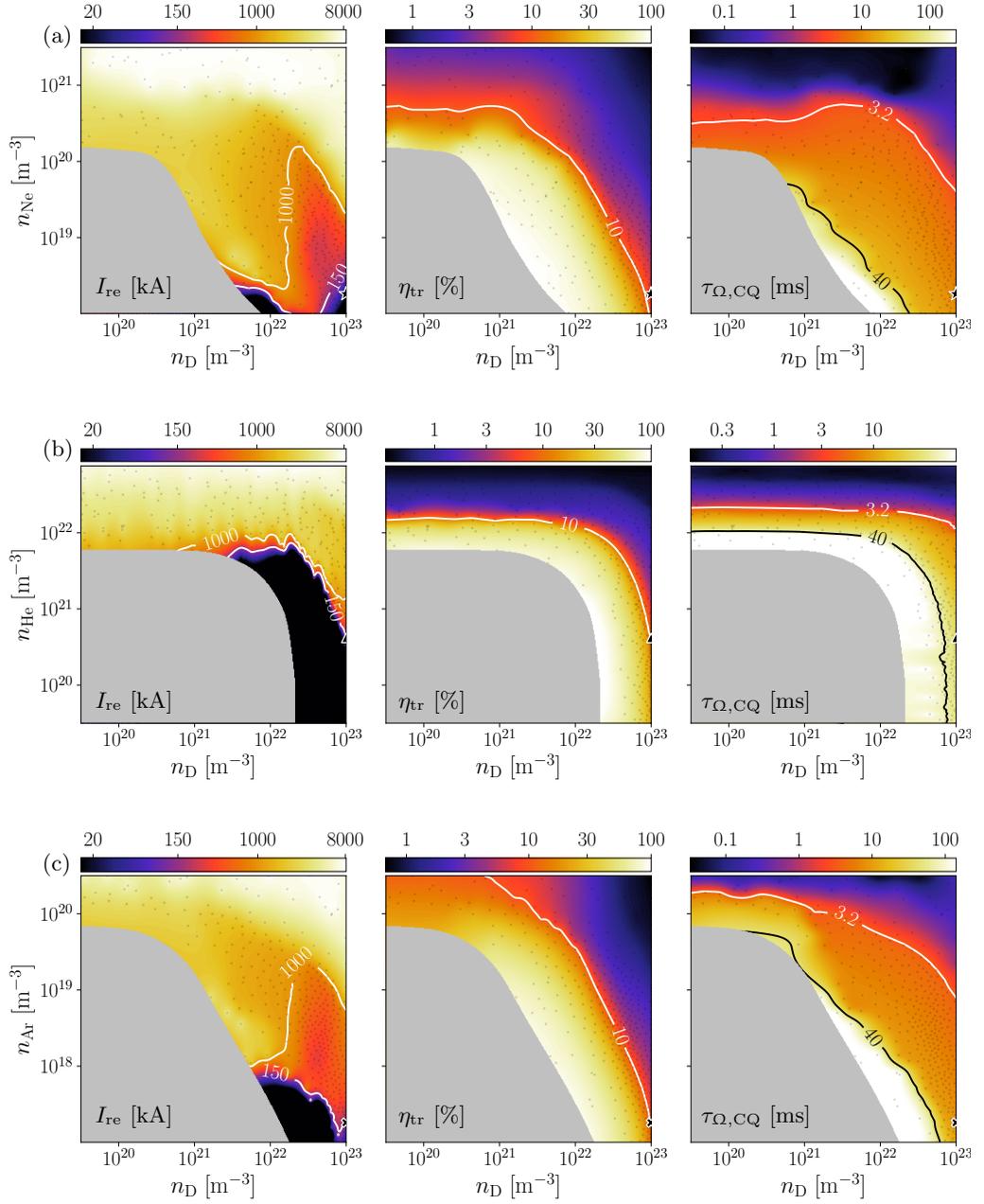
\begin{figure}
    \centering
    \begin{subfigure}[t]{13.44cm}
        \begin{overpic}[width=13.44cm]{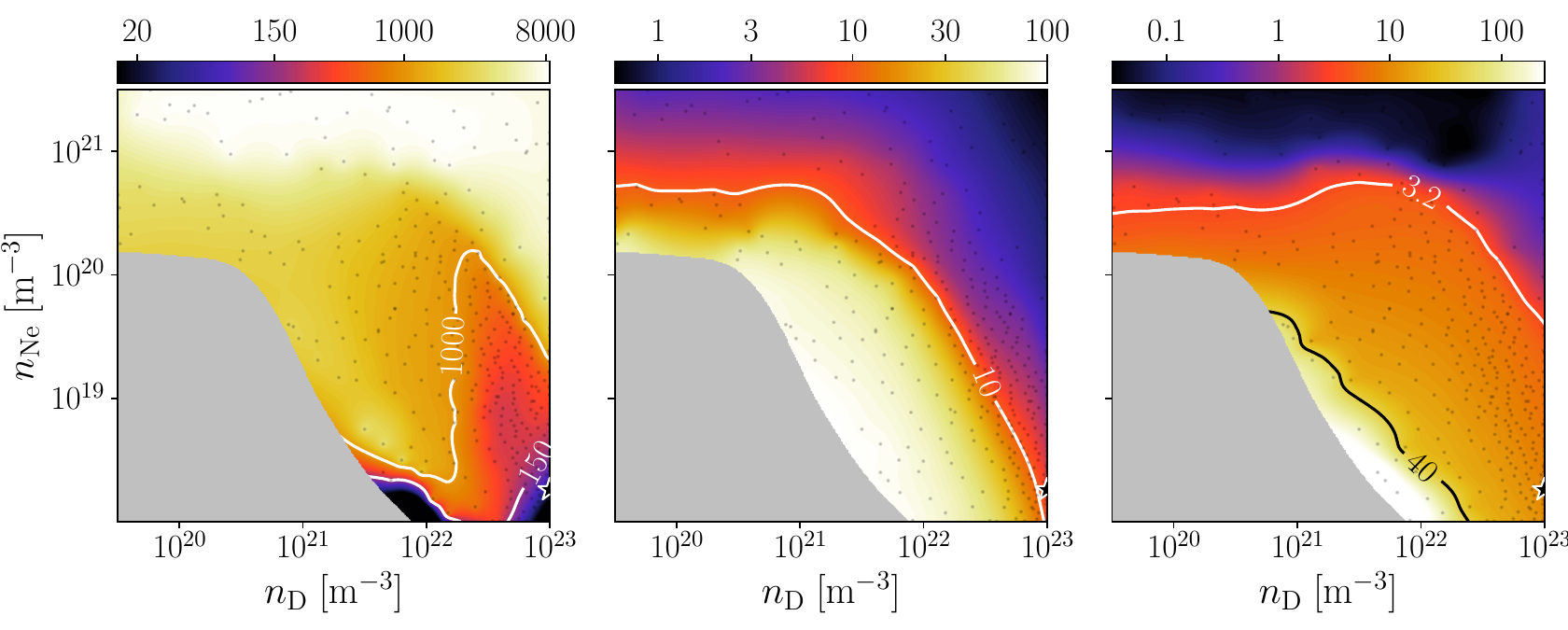}
            \put(3.5, 35){(a)}
            \put(9, 9){$\Ire$ [kA]}
            \put(40.5, 9){$\etac$ [\%]}
            \put(72, 9){$\taucq$ [ms]}
        \end{overpic} 
        \phantomcaption
        \label{fig:comps_Ne}
    \end{subfigure}
    \begin{subfigure}[t]{13.44cm}
        \begin{overpic}[width=13.44cm]{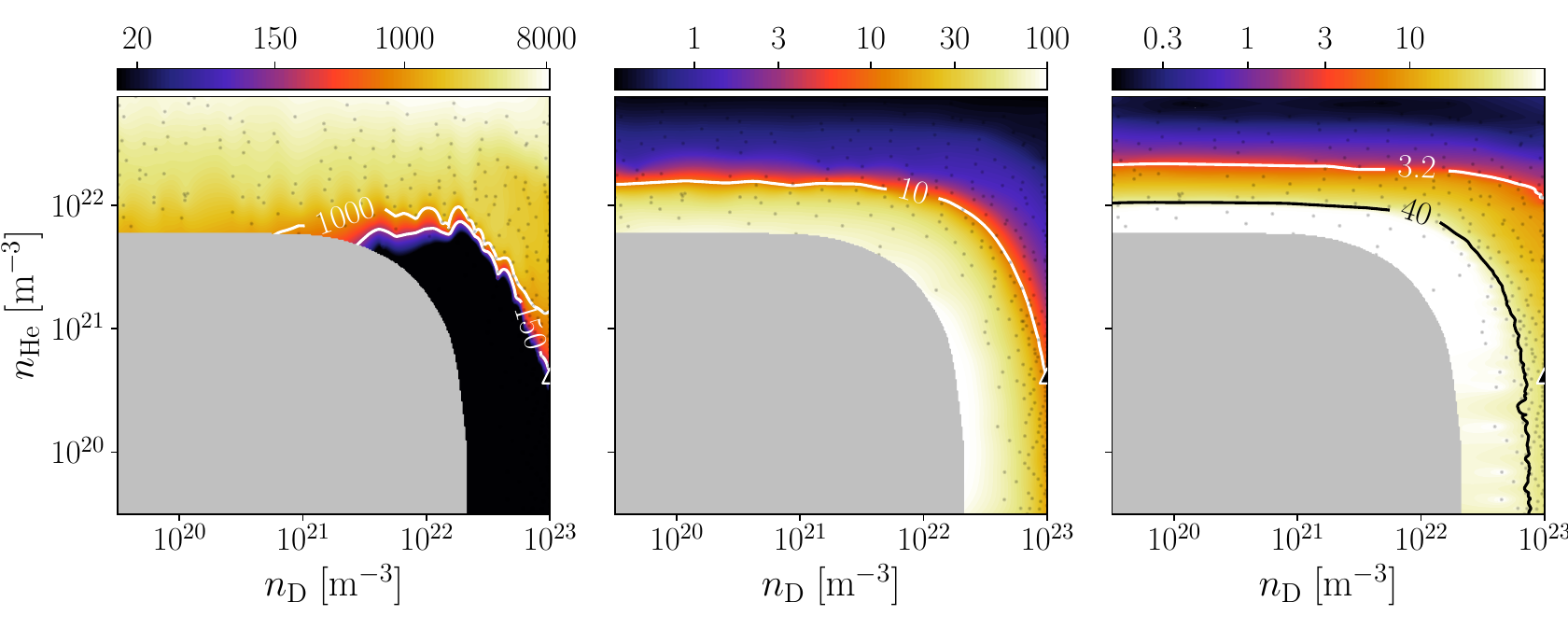}
            \put(3.5, 35){(b)}
            \put(9, 9){$\Ire$ [kA]}
            \put(40.5, 9){$\etac$ [\%]}
            \put(72, 9){$\taucq$ [ms]}
        \end{overpic} 
        \phantomcaption
        \label{fig:comps_He}
    \end{subfigure}
    \begin{subfigure}[t]{13.44cm}
        \begin{overpic}[width=13.44cm]{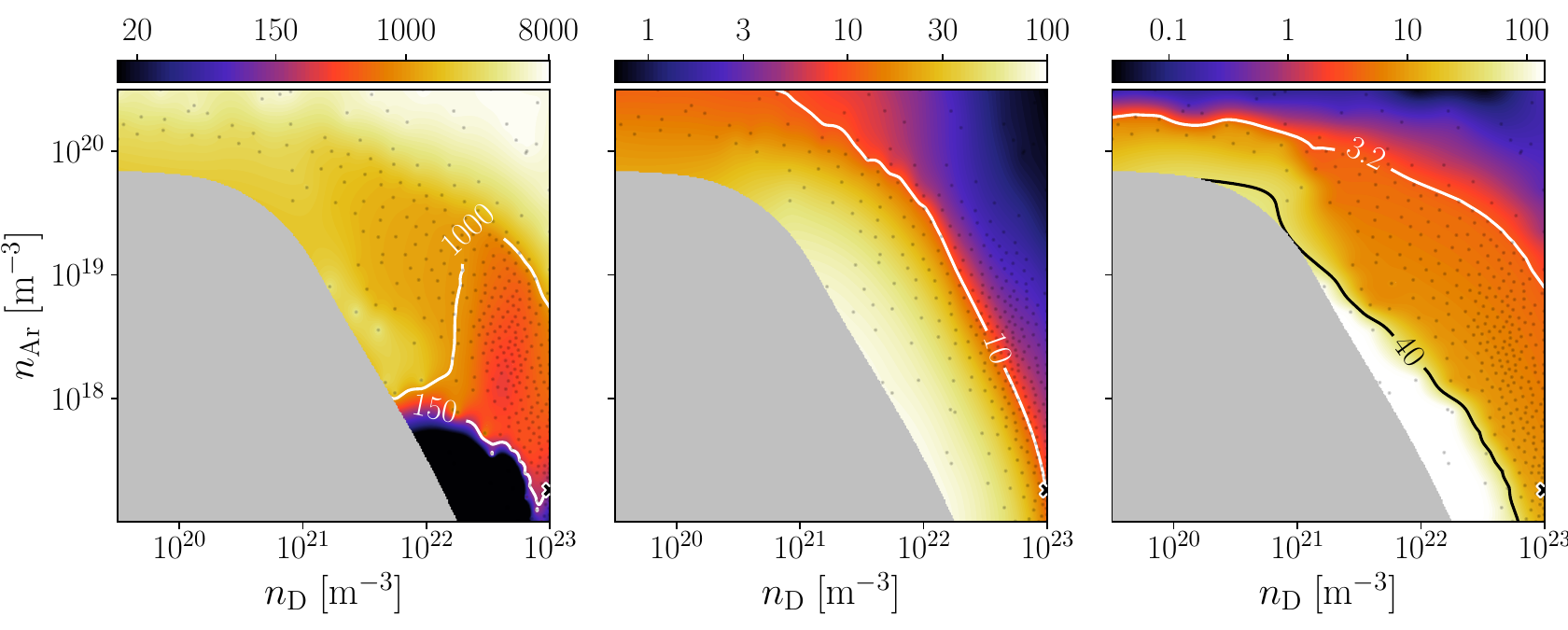}
            \put(3.5, 35){(c)}
            \put(9, 9){$\Ire$ [kA]}
            \put(40.5, 9){$\etac$ [\%]}
            \put(72, 9){$\taucq$ [ms]}
        \end{overpic} 
        \phantomcaption
        \label{fig:comps_Ar}
    \end{subfigure}
    \caption{Logarithmic contour plots of the figure of merit estimates from the optimizations of 
    \deuteriumtritium{} operation with MMI of \deuterium{} in combination with (a) \neon, (b) \helium{} and 
    (c) \argon. 
    }
    \label{fig:comps}
\end{figure}

Here, we present more detailed plots of the figure of merit landscapes in the injected density parameter 
space. 
In figure~\ref{fig:compsNA}, the figure of merit landscapes obtained from the optimization 
of \deuterium{} operation are presented, both with (figure~\ref{fig:comps_D_C}) and without 
(figure~\ref{fig:comps_D}) RE generation from DD-induced Compton scattering. 
The figure of merit landscapes obtained from the optimization of \deuteriumtritium{} operation are 
presented in figure~\ref{fig:comps}, for MMI of \neon{} (figure~\ref{fig:comps_Ne}), \helium{} 
(figure~\ref{fig:comps_He}) and \argon{} (\ref{fig:comps_Ar}).

\begin{figure}
    \centering
    \begin{subfigure}[t]{13.44cm}
        \begin{overpic}[width=13.44cm]{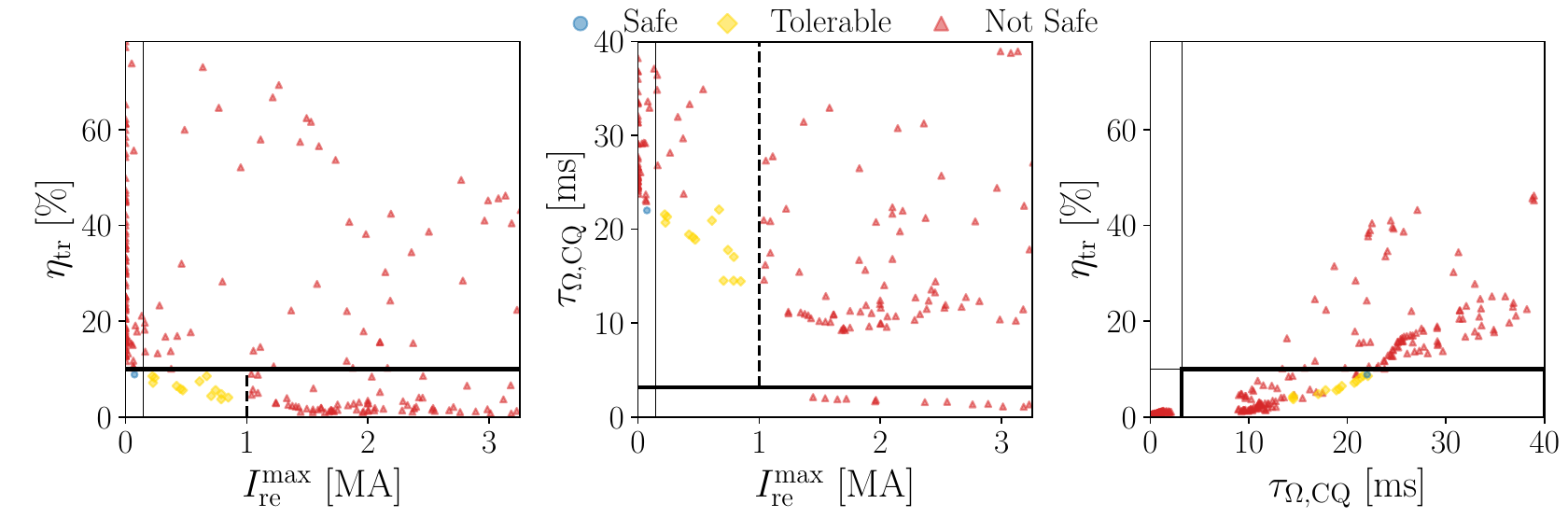}
            \put(0, 29){(a)}
        \end{overpic}
        \phantomcaption
        \label{fig:cloud_He}
    \end{subfigure}
    \begin{subfigure}[t]{13.44cm}
        \begin{overpic}[width=13.44cm]{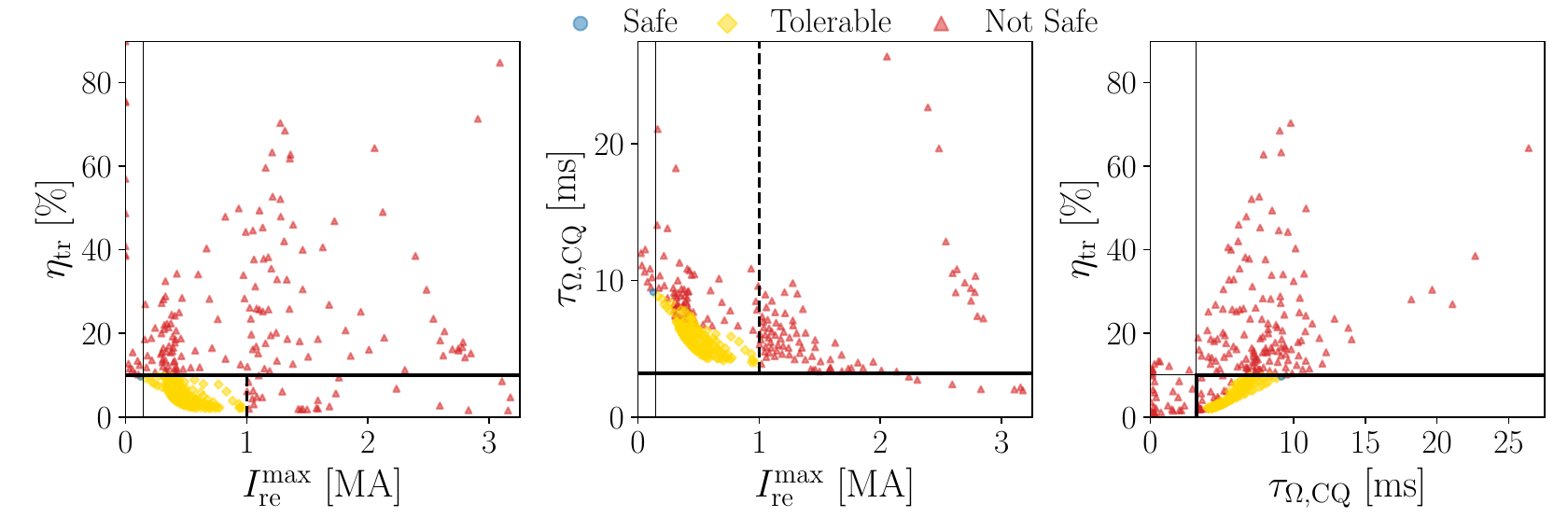}
            \put(0, 29){(b)}
        \end{overpic} 
        \phantomcaption
        \label{fig:cloud_Ar}
    \end{subfigure}
    \caption{Projections of the simulation dataset to all the two-dimensional subspaces of the figure of 
    merit space $(\Ire, \etac, \taucq)$ from the (a) \helium{} and (b) \argon{} MMI optimization.
    }
    \label{fig:cloudsactivated_app}
\end{figure}

The trade-off between the runaway current, transported fraction of the heat loss and CQ time for the 
optimization dataset from \deuteriumtritium{} operation with MMI of \helium{} and \argon{} are plotted in 
figures \ref{fig:cloud_He} and \ref{fig:cloud_Ar}, respectively.

\clearpage
\bibliographystyle{jpp}
\bibliography{ref}

\end{document}